\documentclass[11pt,a4paper]{article}
\pdfoutput=1

\usepackage{amsmath,amssymb,tikz}
\usepackage{graphicx}
 \allowdisplaybreaks
\def\be{\begin{equation}}
\def\ee{\end{equation}}
\def\ba#1\ea{\begin{align}#1\end{align}}
\def\bg#1\eg{\begin{gather}#1\end{gather}}
\def\bm#1\em{\begin{multline}#1\end{multline}}
\def\bmd#1\emd{\begin{multlined}#1\end{multlined}}

\def\({\left(}
\def\){\right)}
\def\[{\left[}
\def\]{\right]}

\def \be {\begin{equation}}
\def \ee {\end{equation}}
\def \ba {\begin{array}}
\def \ea {\end{array}}
\def \bea{\begin{eqnarray}}
\def \eea{\end{eqnarray}}

\def\bea{\begin{eqnarray}}
\def\eea{\end{eqnarray}}

\newcommand{\bit}{\begin{itemize}}  \newcommand{\eit}{\end{itemize}}
\newcommand{\ben}{\begin{enumerate}}  \newcommand{\een}{\end{enumerate}}

\long\def\symbolfootnote[#1]#2{\begingroup%
\def\thefootnote{\fnsymbol{footnote}}\footnote[#1]{#2}\endgroup}

\newcommand{\sysu}{{\it School of Physics and Astronomy, Sun Yat-Sen University, 2 Daxue Road, Zhuhai 519082, China}}

\begin{document}
\thispagestyle{empty}
\begin{center}

~\vspace{20pt}

{\Large\bf An Exact Construction of Codimension two Holography}

\vspace{25pt}

Rong-Xin Miao ${}$\symbolfootnote[1]{Email:~\sf
  miaorx@mail.sysu.edu.cn}

\vspace{10pt}${}$\sysu

\vspace{2cm}

\begin{abstract}
Recently, a codimension two holography called wedge holography is proposed as a generalization of AdS/CFT.  It is conjectured that a gravitational theory in $d+1$ dimensional wedge spacetime is dual to a $d-1$ dimensional CFT on the corner of the wedge. 
In this paper, we give an exact construction of the gravitational solutions for wedge holography from the ones in AdS/CFT. By applying this construction, we prove the equivalence between wedge holography and AdS/CFT for vacuum Einstein gravity, by showing that the classical gravitational action and thus the CFT partition function in large N limit are the same for the two theories. The equivalence to AdS/CFT can be regarded as a ``proof'' of wedge holography in a certain sense.  As an application of this powerful equivalence, we derive easily the holographic Weyl anomaly, holographic Entanglement/R\'enyi entropy and correlation functions for wedge holography. Besides, we discuss the general solutions of wedge holography and argue that they correspond to the AdS/CFT with suitable matter fields. Interestingly, we notice that the intrinsic Ricci scalar on the brane is always a constant, which depends on the tension.  Finally, we generalize the discussions to dS/CFT and flat space holography.  Remarkably, we find that AdS/CFT, dS/CFT and flat space holography can be unified in the framework of codimension two holography in asymptotically AdS.  Different dualities are distinguished by different types of spacetimes on the brane. 
\end{abstract}

\end{center}

\newpage
\setcounter{footnote}{0}
\setcounter{page}{1}

\tableofcontents

\section{Introduction}

The holographic principle \cite{tHooft:1993dmi,Susskind:1994vu} reveals a deep connection between a higher dimensional gravity theory and a lower dimensional quantum field theory (QFT).  As an exact realization of holographic principle, the AdS/CFT correspondence \cite{Maldacena:1997re,Gubser:1998bc,Witten:1998qj} conjectures that the quantum gravity theory in an asymptotically anti-de Sitter space (AdS) is dual to the conformal field theory (CFT) on the boundary.  It provides a powerful method to study the non-perturbative phenomena in gauge theories, and has a wide applications in QFT \cite{Sakai:2004cn,Erlich:2005qh,Sakai:2005yt}, quantum information \cite{Rangamani:2016dms} and condensed matter physics \cite{Hartnoll:2009sz}.

There are many interesting generalizations of AdS/CFT, such as dS/CFT \cite{Strominger:2001pn,Maldacena:2002vr,Alishahiha:2004md,Alishahiha:2005dj,Dong:2018cuv}, Kerr/CFT \cite{Guica:2008mu,Castro:2010fd}, flat space holography \cite{Bagchi:2010zz,Bagchi:2016bcd}, brane world holography \cite{Randall:1999ee,Randall:1999vf,Karch:2000ct}, surface/state correspondence \cite{Miyaji:2015yva,Takayanagi:2018pml} and AdS/BCFT \cite{Takayanagi:2011zk,Fujita:2011fp,Nozaki:2012qd,Miao:2018qkc,Miao:2017gyt,Chu:2017aab}. Recently,  a codimension two holography, called wedge holography, is proposed by \cite{Akal:2020wfl} between the gravitational theory in a $d+1$ dimensional wedge spacetime and the $d-1$ dimensional CFT on the corner of the wedge. See also \cite{Bousso:2020kmy} for a similar proposal of codimension two holography. The geometry of wedge holography is showed in figure 1 (left). It is conjectured that the following dualities hold
\begin{eqnarray}\label{Wedgeholography}
\text{Classical gravity on wedge} \ W_{d+1}  &\simeq& \text{(Quantum) gravity on two AdS}_d  (Q_1\cup Q_2)  \nonumber\\
&\simeq &\text{CFT}_{d-1} \ \text{on}\  \Sigma \nonumber
\end{eqnarray}
where the first equivalence is due to the brane world holography  \cite{Randall:1999ee,Randall:1999vf,Karch:2000ct} and the second equivalence originates from AdS/CFT.  It is closely related to the so-called doubly holographic model, which is recently developed for the resolution of information paradox, in particular, for recovering Page curve of Hawking radiation \cite{Penington:2019npb,Almheiri:2019psf,Almheiri:2019hni}.  The main difference from the wedge holography is that the double holography focus on a larger region of AdS instead of only the wedge. See also \cite{Rozali:2019day,Chen:2019uhq,Almheiri:2019psy,Kusuki:2019hcg,Balasubramanian:2020hfs,Sully:2020pza,Geng:2020qvw,Chen:2020uac,Dong:2020uxp,Arias:2019zug,Arias:2019pzy,Geng:2020kxh} for related topics. 
The gravitational action of wedge holography in Euclidean signature is given by  \cite{Akal:2020wfl} 
\begin{eqnarray}\label{action}
  I=-\frac{1}{16\pi G_N}\int_N \sqrt{g} (R-2\Lambda)
  -\frac{1}{8\pi G_N}\int_{Q_1 \cup Q_2} \sqrt{h} (K-T),
\end{eqnarray}
where $N$ denotes $d+1$ dimensional wedge space, $Q_1$ and $Q_2$ denote two $d$ dimensional branes. \cite{Akal:2020wfl}  propose to impose Neumann boundary condition (NBC) on the two end-of-world branes
\begin{eqnarray}\label{NBC}
K^i_j -(K-T) h^i_j=0,
\end{eqnarray}
where $K_{ij}$ are the extrinsic curvatures, $h_{ij}$ are the induced metrics on the brane and $T$ is the brane tension. Notice that the wedge holography can be regarded as a limit of AdS/BCFT \cite{Takayanagi:2011zk} with vanishing width of strip \cite{Akal:2020wfl}. See figure 1 (right) for example. 

\begin{figure}[t]
\centering
\includegraphics[width=6cm]{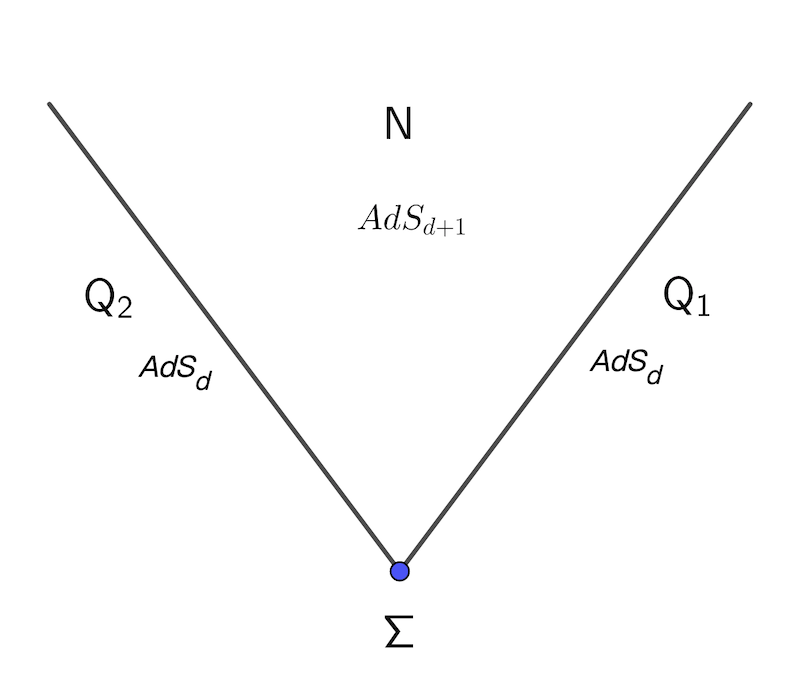}
\includegraphics[width=6.8cm]{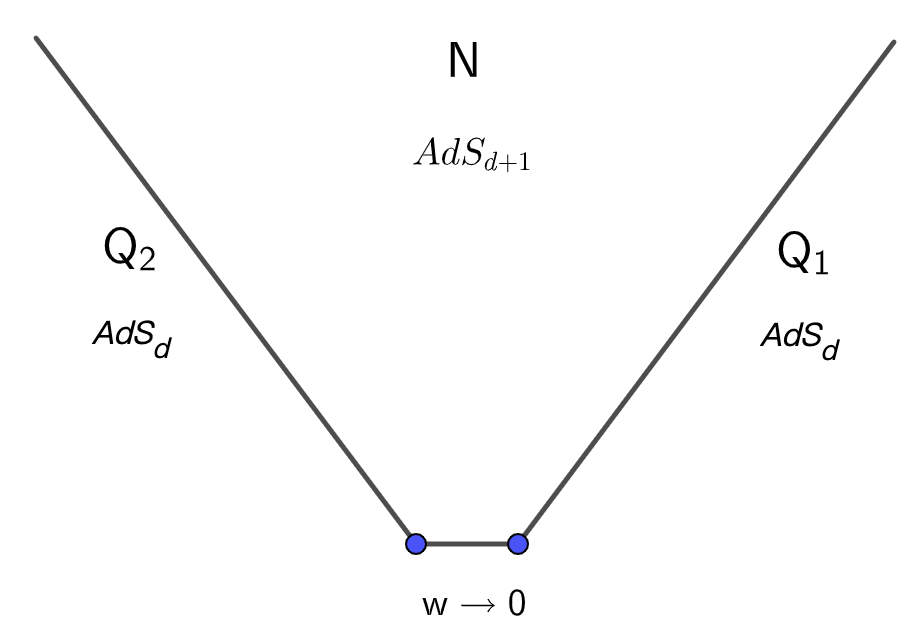}
\caption{(left) Geometry of wedge holography;\  \ (right) Wedge holography from AdS/BCFT}
\label{wedge}
\end{figure}

In \cite{Akal:2020wfl}, the authors test wedge holography by studying free energy, Weyl anomaly of 2d CFT, entanglement entropy, two point functions of scalar operators and so on. For simplicity, they mainly focus on (locally) AdS in the bulk N and on the branes Q. 

In this paper, we provide more supports for wedge holography. For simplicity, we focus on the codimension two correspondence, i.e.,  $ \text{classical gravity on wedge} \ W_{d+1}  \simeq  \text{CFT}_{d-1}$.  We find an exact map from the solution to vacuum Einstein equations in $\text{AdS}_{d}/\text{CFT}_{d-1}$ to the solution in wedge holography $\text{AdSW}_{d+1}/\text{CFT}_{d-1}$. By using this map, we prove that wedge holography with this novel class of solutions is equivalent to AdS/CFT with vacuum Einstein gravity.  Assuming that AdS/CFT is correct, which is widely accepted, this equivalence is actually a proof of wedge holography in a certain sense. Now most of the results of AdS/CFT apply directly to wedge holography. For examples, we calculate holographic Weyl anomaly, holographic R\'enyi  entropy and correlation functions for wedge holography and find that them all take the correct forms.  We also discuss the more general solutions in wedge holography and argue that they are equivalent to AdS/CFT with matter fields. Finally, we generalize the discussions to de Sitter space (dS) and flat space (Minkowski) on the branes.  Remarkably, we find that AdS/CFT, dS/CFT and flat space holography can be unified in the framework of codimension two holography in asymptotically AdS. 

It should be stressed that, mathematically, the maps from the solutions to Einstein equations in $d$ dimensions to the ones in $d+1$ dimensions (theorems I-VI of this paper) are not new results. In fact, they are well-known results in the context of the brane world \footnote{We thank the referee to point out this fact to us.}. See  \cite{Emparan:1999pm,Park:2001jh} for example. The main result of this paper is that it provides new perspectives of wedge holography by bring theorems I-VI into this context, and it provides more evidences for wedge holography by carefully studying physical quantities such as Weyl anomaly, R\'enyi entropy and two point functions of the energy stress tensor.

The paper is organized as follows. 
In section 2, we give an exact construction of the gravitational solutions in wedge holography from the ones in AdS/CFT. By using this construction, we prove that wedge holography is equivalent to AdS/CFT, at least for a novel class of solutions. In section 3, we test wedge holography by studying the holographic Weyl anomaly, holographic R\'enyi entropy and correlation functions. In section 4, we discuss the general solutions in codimension two holography for wedges.  In section 5, we generalize our results to dS/CFT and flat space holography. Finally, we conclude with some open problems in section 6.

\section{Construction of codimension two holography}

In this section, we construct the gravitational solution for wedge holography (or $\text{AdS}_{d+1}/\text{BCFT}_{d}$ generally) in $d+1$ dimensions from the one in $\text{AdS}_d/\text{CFT}_{d-1}$ in d dimensions.  
By using this construction, we prove that the gravitational action of wedge holography is equivalent  to that of AdS/CFT, which means
\begin{eqnarray}\label{actionequal}
\text{classical gravity on wedge} \ W_{d+1}  \simeq \text{classical gravity in AdS}_d .
\end{eqnarray}
Assuming that $\text{AdS}_d/\text{CFT}_{d-1}$ holds
\begin{eqnarray}\label{AdSCFT}
\text{classical gravity in AdS}_d   \simeq \text{CFT}_{d-1},
\end{eqnarray}
we conclude immediately that  the wedge holography also holds 
\begin{eqnarray}\label{AdSWCFT}
 \text{classical gravity on wedge} \ W_{d+1}  \simeq  \text{CFT}_{d-1}. 
 \end{eqnarray}
Note that following \cite{Akal:2020wfl} we focus on the classical gravity in this paper. In general, AdS/CFT and wedge holography also apply to quantum gravity.

\subsection{Exact constructions of solutions}

Let us start with the following ansatz of the metric
\begin{eqnarray}\label{keysolution}
ds^2=g_{\mu\nu}dx^{\mu}dx^{\nu}=dx^2+\cosh^2(x) \bar{h}_{ij}(y) dy^i dy^j,
\end{eqnarray}
where $x^{\mu}=(x,y^i)$, $g_{\mu\nu}$ and $\bar{h}_{ij}$  are the metrics in $d+1$ dimensions and $d$ dimensions, respectively.  Note that $\bar{h}_{ij}(y)$ are independent of the normal coordinate x. Without loss of generality, we set the AdS radius $L=1$ in this paper.  When $\bar{h}_{ij}$ is the $\text{AdS}_d$ metric, $g_{\mu\nu}$ turns out to be  $\text{AdS}_{d+1}$ metric \cite{Karch:2000ct,Akal:2020wfl}. 
Actually $\bar{h}_{ij}$ can be relaxed to be any metric obeying Einstein equations in the framework of codimension two holography, or generally, the AdS/BCFT \cite{Takayanagi:2011zk}.  Let us present the above statements in two theorems. 
Let us emphasize again that the following theorems are not new results. See \cite{Emparan:1999pm,Park:2001jh} for similar results in the context of the brane world.  What is new in this paper is that, it provides new perspectives of wedge holography by bring these theorems into this context, and it provides more evidences for wedge holography. See section 3 for examples. 

{\bf Theorem I} \\
The metric (\ref{keysolution}) is a solution to Einstein equation with a negative cosmological constant in $d+1$ dimensions 
\begin{eqnarray}\label{EOMg}
R_{\mu\nu}-\frac{R}{2}g_{\mu\nu}=\frac{d(d-1)}{2} g_{\mu\nu}
\end{eqnarray}
provided that $\bar{h}_{ij}$ obey Einstein equation with a negative cosmological constant in d dimensions
\begin{eqnarray}\label{EOMh}
R_{\bar{h}\ ij}-\frac{R_{\bar{h}}}{2}\bar{h}_{ij}=\frac{(d-1)(d-2)}{2} \bar{h}_{ij}.
\end{eqnarray}
Here $R_{\bar{h}\ ij}$ denote the curvatures with respect to the metric $\bar{h}_{ij}$ and we have set the AdS radius $L=1$ for simplicity.

{\bf Proof I} \\
The proof of theorem I is straightforward.  From (\ref{EOMh}), we get
\begin{eqnarray}\label{Rhij}
R_{\bar{h}\ ij}=-(d-1) \bar{h}_{ij}=-(d-1)\text{sech}^2(x) g_{ij}
\end{eqnarray}
Substituting the above equation into (\ref{curvature2}) of the appendix, after some simple calculations we obtain
\begin{eqnarray}\label{Rgij}
R_{xx}=-d g_{xx}, \  R_{xi}=0, \  R_{ij}=-d g_{ij},
\end{eqnarray}
which are the equivalent expressions of Einstein equations (\ref{EOMg}), i.e., $R_{\mu\nu}=-d g_{\mu\nu}$. Now we finish the proof that the metric (\ref{keysolution}) is a solution to Einstein equation in $d+1$ dimensions as long as $\bar{h}_{ij}$ is a solution to  Einstein equation in d dimension.

{\bf Theorem II} \\
The metric (\ref{keysolution}) satisfies the NBC (\ref{NBC}) on the branes located at $x=\pm \rho$. Here we have re-parameterized the tension as $T=(d-1) \tanh\rho$ in (\ref{NBC}).

{\bf Proof II} \\
Note that the metric (\ref{keysolution}) is written in Gauss normal coordinates. For this special kind of coordinates, the extrinsic curvatures take simple form 
\begin{eqnarray}\label{extrinsic1}
&&K_{ij}=\pm \frac{1}{2} (\partial_x g_{ij})|_{x=\pm\rho}=\tanh\rho \ g_{ij}|_{x=\pm\rho},\\
&&K=g^{ij}K_{ij}= d \tanh\rho \label{extrinsic2}
\end{eqnarray}
which indeed obeys NBC (\ref{NBC}). Please see the appendix for more discussions of extrinsic curvatures.   Now we finish the proof.

Some comments are in order.
 {\bf 1}  As mentioned above, theorem I and theorem II are already known in the context of the brane world  \cite{Emparan:1999pm,Park:2001jh}.  Mathematically, theorem I and theorem II of this paper are not new. Physically, as we will show in the rest of this paper, they bring new perspectives for wedge holography and provide a useful tool to study wedge holography. For example, by applying these theorems, one can prove the  equivalence between Wedge holography and AdS/CFT for a novel class of solutions. By using this equivalence, one can derive directly Weyl anomaly, R\'enyi entropy and so on from the approach of AdS/CFT. Besides, these theorems give a good start point to explicitly examine the physics on the d-dimensional branes for wedge holography. {\bf 2} According to theorem I and theorem II, it is clear that the metric (\ref{keysolution}) is a solution to AdS/BCFT \cite{Takayanagi:2011zk} with range $-\infty \le x\le \rho$, and a solution to wedge holography \cite{Akal:2020wfl} with range $-\rho \le x\le \rho$ \footnote{Actually, the wedge holography can be defined in a more general region $-\rho_1 \le x\le \rho_2$ with $\rho_1 \ne \rho_2$. Now the tensions of the two branes are different. However, nothing goes wrong and (\ref{keysolution}) is still a solution for this general case. }. Note that the wedge holography can be regarded as a special limit of AdS/BCFT \cite{Akal:2020wfl}.  {\bf 3}  The metric  (\ref{keysolution}) defines a map from the gravitational solutions in d dimensions to those in one dimension higher. By using this map,  one can obtain non-trivial exact black hole solutions in AdS/BCFT. We leave a careful study of the property of black holes in AdS/BCFT in \cite{ChuandMiao}.  {\bf 4} For simplicity, we focus on vacuum Einstein gravity in this paper. 
The discussions can be generalized to the case with matter fields \cite{HuandMiao}.  {\bf 5} Of course,  (\ref{keysolution}) is not the most general solution to Einstein equations in $d+1$ dimensions. That is because the solution space in higher dimensions is larger than the one in lower dimensions.  As we will shown below, the solution (\ref{keysolution}) is of special interest due to the evident relation to AdS/CFT \cite{Maldacena:1997re}. We leave the discussions of other kinds of solutions in section 4 and section 5.

\subsection{Equivalence to AdS/CFT}

Now let us prove that the wedge holography $\text{AdSW}_{d+1}/\text{CFT}_{d-1}$ with the solution (\ref{keysolution}) is equivalent to $\text{AdS}_d/\text{CFT}_{d-1}$ with vacuum Einstein gravity.  The $\text{AdSW}_{d+1}/\text{CFT}_{d-1}$ claims that the partition function of $\text{CFT}_{d-1}$ in large N limit is given by the classical gravitational action in $d+1$ dimensional asymptotically AdS spacetime with a wedge
\begin{eqnarray}\label{AdSWCFT}
Z_{\text{CFT}_{d-1}}=e^{-I_{\text{AdSW}_{d+1}}}.
\end{eqnarray}
Similarly, the $\text{AdS}_{d}/\text{CFT}_{d-1}$ assumes that
\begin{eqnarray}\label{AdSWCFT}
Z_{\text{CFT}_{d-1}}=e^{-I_{\text{AdS}_d}}.
\end{eqnarray}
To prove the equivalence between wedge holography and AdS/CFT, we only need to prove that $I_{\text{AdSW}_{d+1}}=I_{\text{AdS}_d}$. Since the solution  (\ref{keysolution}) is symmetrical for $x$, for simplicity we focus on half of the wedge spacetime $0\le x \le \rho$ below. Please keep in mind that the total gravitational action is double of the result below. 

Substituting the metric (\ref{keysolution}) into the action (\ref{action}) and applying NBC (\ref{NBC}) together with the formula (\ref{curvature3}), we derive
\begin{eqnarray}\label{AdSWaction}
I_{\text{AdSW}_{d+1}}
&=&\frac{1}{16\pi G_N}\int_0^{\rho} \cosh^{d}(x)dx \int_{Q_1} \sqrt{\bar{h}} \Big{(} R_{\bar{h}} \text{sech}^2(x)- d\left( 2+(d-1) \tanh^2x\right)+d(d-1) \Big{)}\nonumber\\
&+&\frac{1}{8\pi G_N}\int_{Q_1} \sqrt{\bar{h}} \cosh^d\rho \tanh\rho \nonumber\\
&=& \frac{1}{16\pi G_N}\int_0^{\rho} \cosh^{d-2}(x)dx \int_{Q_1} \sqrt{\bar{h}} \Big{(} R_{\bar{h}} +(d-1)(d-2) \Big{)} \nonumber\\
&=& \frac{1}{16\pi G^{(d)}_N}\int_{Q_1} \sqrt{\bar{h}} \Big{(} R_{\bar{h}} +(d-1)(d-2) \Big{)}=I_{\text{AdS}_d}\ ,
\end{eqnarray}
which is equal to the gravitational action $I_{\text{AdS}_d}$ with Newton's constant given by
\begin{eqnarray}\label{Newton's constant}
\frac{1}{G^{(d)}_N}=\frac{1}{ G_N}\int_0^{\rho} \cosh^{d-2}(x)dx.
\end{eqnarray}
Note that we take $\bar{h}_{ij}$ off-shell in the above derivations. In other words, we do not need to impose Einstein equations 
on the brane $Q$ (\ref{EOMh}). In fact,  Einstein equations (\ref{EOMh}) on $Q$ can be derived by the variation of $I_{\text{AdSW}_{d+1}}$ with respect to $\bar{h}_{ij}$.  Besides, we have used $-2\Lambda=d(d-1)$, $K-T=\tanh\rho$ and the following formula
\begin{eqnarray}\label{integralformula}
\int_0^{\rho} \cosh ^{d-2}x \left(d-1-d \cosh ^2x \right) dx=- \cosh^d\rho \tanh\rho .
\end{eqnarray}

In the above discussions, we focus on the non-renormalized action (\ref{AdSWaction}), which is divergent generally. To get the finite result, one can perform the holographic renormalization \cite{Balasubramanian:1999re,deHaro:2000vlm} by adding suitable counterterms on $\Sigma$.  Following \cite{Balasubramanian:1999re,deHaro:2000vlm}, we choose the following counterterms on the corner of the wedge $\Sigma$
\begin{eqnarray}\label{counterterms}
I_C=\frac{1}{16\pi G^{(d)}_N} \int_{\Sigma} \sqrt{\sigma} \left( 2 K_{\Sigma} +2(2-d)-\frac{1}{d-3} R_{\Sigma}+...\right),
\end{eqnarray}
which makes the equivalence
\begin{eqnarray}\label{keyequivalence}
I_{\text{AdSW}_{d+1}}+I_C=I_{\text{AdS}_d}+I_C
\end{eqnarray}
still holds after renormalization. Note that, in order to have a well-defined action variation, one need to add a Hayward term \cite{Hayward:1993my,Brill:1994mb} at the corner of the wedge
\begin{eqnarray}\label{Haywardterm}
I_H=\frac{1}{8\pi G_N} \int_{\Sigma} \sqrt{\sigma} (\Theta-\pi),
\end{eqnarray}
where $\Theta$ is the angle between two branes.  As a constant, the Hayward term (\ref{Haywardterm}) can be absorbed into the counterterms (\ref{counterterms}) of holographic renormalization. 
Now we finish the proof of the statement that $\text{AdSW}_{d+1}/\text{CFT}_{d-1}$ with the solution (\ref{keysolution}) is equivalent to $\text{AdS}_d/\text{CFT}_{d-1}$, at least at the classical level for gravity, or equivalently, in the large N limit for CFTs. 

Some comments are in order. 
{\bf 1}. The equivalence (\ref{AdSWaction}) between gravitational action follows naturally from the map (\ref{keysolution}) between gravitational solutions for wedge holography and AdS/CFT. The interesting point is that, to derive (\ref{AdSWaction}), we can take $\bar{h}_{ij}$ off-shell on the brane. In other words, we do not need to use theorem I in order to derive (\ref{AdSWaction}).   {\bf 2}. The equivalence (\ref{keyequivalence}) is quite powerful which enables us to derive many interesting physical quantities such as Entanglement/R\'enyi entropy for wedge holography directly following the approach of AdS/CFT. See sect.3 for examples. {\bf 3}. Assuming that AdS/CFT holds which is widely accepted, the equivalence (\ref{keyequivalence}) is actually a proof of the wedge holography in a certain sense. 
{\bf 4}. Although we only prove the equivalence (\ref{keyequivalence}) for vacuum Einstein gravity, the equivalence (\ref{keyequivalence}) is expected to hold generally with matter fields.  We leave the generalization to matters in future works. {\bf 5}. As we have mentioned in above section,  the solution (\ref{keyequivalence}) is not the most general solution to vacuum Einstein equations in $d+1$ dimensions. There are other kinds of solutions, which may correspond to different kinds of dualities. For example,  the wedge holography can also be equivalent to dS/CFT if the spacetime on the brane is asymptotically dS instead of AdS. Please see section 5 for more details.  {\bf 6}. Even for the case with asymptotically AdS branes, (\ref{keyequivalence}) is not the most general solution. See sect. 4 for the discussion of more general solutions. We argue that they are equivalent to AdS/CFT with suitable matter fields. That is reasonable since there are Kaluza-Klein modes after the dimensional reduction in the $x$ direction. 

\section{Aspects of wedge holography}

Since wedge holography with the solution (\ref{keysolution}) is equivalent to AdS/CFT with vacuum Einstein gravity,  many novel results of AdS/CFT can be recovered in wedge holography. In this section, we study holographic Weyl anomaly, holographic Entanglement/R\'enyi entropy and holographic correlation functions and provide more supports for the wedge holography. It is interesting that holographic g-theorem for BCFTs becomes holographic c-theorem for CFTs in the framework of wedge holography. 

\subsection{Holographic Weyl anomaly}

Weyl anomaly measures the violation of scaling symmetry of conformal field theory (CFT) due to quantum effects.  For the CFT without boundaries, Weyl anomaly appears only in even dimensions.  In general, it takes the following form \cite{Weylanomaly}
\begin{eqnarray}\label{Weylanomaly}
\mathcal{A}=\int_{\Sigma} dx^{2p} \sqrt{\sigma} [ \sum_n B_n I_n  -2(-1)^{p} A\  E_{2p} ]
\end{eqnarray}
where $2p=d-1$ is even, $A, B_n$ are central charges, $I_n$ are the Weyl invariant terms constructed from curvatures and their covariant derivatives, and $E_{2p}$ is the Euler density defined by
\begin{eqnarray}\label{E2p}
E_{2p}(R)=\frac{1}{(8\pi)^p\Gamma(p+1)} \delta^{i_1 i_2...i_{2p-1}i_{2p}}_{j_1 j_2...j_{2p-1}j_{2p}} R^{j_1 j_2}_{\ \ \ \ i_1i_2}... R^{j_{2p-1} j_{2p}}_{\ \ \ \ \ \ \ \  i_{2p-1}i_{2p}},
\end{eqnarray}
which is normalized so that integrated over a d-dimensional sphere: $\oint_{S^d} \sqrt{\sigma} E_d=2$  \cite{Hung:2011xb}. 
It is conjectured that the central charges related to Euler density obey the c-theorem \cite{Cardy:1988cwa}
\begin{eqnarray}\label{atheorem}
A_{UV}\ge A_{IR}.
\end{eqnarray}
So far the c-theorem is proved only for 2d CFTs \cite{Zamolodchikov:1986gt} and 4d CFTs  \cite{Komargodski:2011vj} . In higher dimensions, there are interesting holographic proofs and generalizations \cite{Myers:2010tj}. See also  \cite{Osborn:1989td,Freedman:1999gp,Girardello:1998pd,Girardello:1999bd,Myers:2010xs} for related works. 

Let us consider some examples of Weyl anomaly. For 2d CFTs and 4d CFTs, the more conventional nomenclature is
\begin{eqnarray}\label{Weylanomaly2d}
&&\mathcal{A}_{2d}=\int_{\Sigma} dx^{2}\sqrt{\sigma} \frac{c_{2d}}{24 \pi} R_{\Sigma},\\
&&\mathcal{A}_{4d}=\int_{\Sigma} dx^{4}\sqrt{\sigma} [\frac{c}{16\pi^2} C_{\Sigma}^{ijkl}C_{\Sigma\ ijkl}-\frac{a}{16\pi^2}(R_{\Sigma}{}^{ijkl}R_{\Sigma}{}_{ijkl}-4R_{\Sigma}{}^{ij}R_{\Sigma}{}_{ij}+R_{\Sigma}^2)  ] ,\label{Weylanomaly4d}
\end{eqnarray}
where $C_{\Sigma\ ijkl}$ are the Weyl tensor on $\Sigma$. 

In the holographic theory, Weyl anomaly can be obtained from the UV logarithmic divergent term of the gravitational action \cite{Henningson:1998gx}.  We assume that the spacetime on the brane is asymptotically AdS
\begin{eqnarray}\label{metricanomaly}
ds^2=dx^2+\cosh^2(x) \frac{dz^2+\sigma_{ij} dy^i dy^j}{z^2},
\end{eqnarray}
where $\sigma_{ij} =\sigma^{(0)}_{ij} + z^2 \sigma^{(1)}_{ij} +...+z^{d-1} ( \sigma^{(d-1)}_{ij}+\lambda^{(d-1)}_{ij}\ln z)$.  Solving Einstein equations (\ref{EOMg}) in $d+1$ dimenions, we get
\begin{eqnarray}\label{g12d}
&& \sigma^{(0)ij}\sigma^{(1)}_{ij}=-\frac{R_{\Sigma}}{2},  \ \ \ \ \ \ \ \ \ \ \ \ \ \ \ \ \ \ \ \ \ \ \ \ \text{for} \ d=3 \\
&& \sigma^{(1)}_{ij}=\frac{-1}{d-3}(R_{\Sigma\ ij}-\frac{R_{\Sigma}}{2(d-2)}\sigma^{(0)}_{ij} ), \ \ \text{for} \  d>3 \label{g1hd}
\end{eqnarray}
Note that $\sigma^{(1)}_{ij}$ can also be obtained  from the asymptotical symmetry of AdS \cite{Imbimbo:1999bj}. Substituting the metric (\ref{metricanomaly}) with (\ref{g12d},\ref{g1hd}) into the action (\ref{action}) and selecting the  UV logarithmic divergent term, we can derive Weyl anomaly (\ref{Weylanomaly2d},\ref{Weylanomaly4d}) for 2d and 4d CFTs with the central charges given by 
\begin{eqnarray}\label{charge2d}
&& c_{2d}=\frac{3}{2 G_N}\int_0^{\rho} \cosh(x)dx=\frac{3}{2 G_N} \sinh(\rho),\\
&& a=c=\frac{\pi}{8 G_N}\int_0^{\rho} \cosh^{3}(x)dx=\frac{\pi}{96 G_N}  (9 \sinh (\rho )+\sinh (3 \rho )) .\label{charge4d}
\end{eqnarray}
Note that we consider only half of the wedge in the above calculations, the total central charges for the whole wedge are double of the results above. Note also that we do not need to know $ \sigma^{(2)}_{ij} $ of order $O(R_{\Sigma}^2)$ in the above derivations for 4d CFTs. That is because the logarithmic terms including $ \sigma^{(2)}_{ij} $ vanish for 4d CFTs in the so-called `off-shell' method \cite{Miao:2013nfa}. 

Let us go on to consider the Weyl anomaly in higher dimensions. For simplicity, we focus on the A-type anomaly $E_{d-1}$. For our purpose, it is sufficient to consider CFTs living on the sphere so that all of the B-type anomaly $I_n$ of (\ref{Weylanomaly}) vanish. Thus we take the the following metric
\begin{eqnarray}\label{metricanomalyhd}
ds^2=dx^2+\cosh^2(x) \frac{dz^2+\frac{(1-z^2)^2}{4}d\Omega_{d-1}^2}{z^2},
\end{eqnarray}
where $d\Omega_{d-1}^2$ denote the volume element of $(d-1)$-dimensional unit sphere. 
Substituting (\ref{metricanomalyhd}) into the action (\ref{action}) and selecting the  UV logarithmic divergent term, we easily get the Weyl anomaly
\begin{eqnarray}\label{Weylanomalygenereld}
\mathcal{A}=(-1)^{\frac{d+1}{2}}\frac{\pi ^{\frac{d-3}{2}}}{2 \Gamma \left(\frac{d-1}{2}\right)}\frac{1}{ G_N}\int_0^{\rho} \cosh^{d-2}(x)dx.
\end{eqnarray}
Comparing (\ref{Weylanomalygenereld}) with (\ref{Weylanomaly}) and noting that $\oint_{S^{d-1}} \sqrt{\sigma} E_{d-1}=2, \ I_n=0$ for spheres, we read off the central charge
\begin{eqnarray}\label{chargegenereld}
A=\frac{\pi ^{\frac{d-3}{2}}}{8 \Gamma \left(\frac{d-1}{2}\right)}\frac{1}{G_N}\int_0^{\rho} \cosh^{d-2}(x)dx.
\end{eqnarray}
From the Weyl anomaly (\ref{Weylanomaly}), one can derive the universal term (log term) of entanglement entropy as \cite{Hung:2011xb}
\begin{eqnarray}\label{EEfromanomaly}
S_{EE}|_{ \ln \frac{1}{\epsilon}}=4(-1)^{(d+1)/2} A,
\end{eqnarray}
when the entangling surface is a sphere.  Later, we will see that (\ref{EEfromanomaly}) agrees with the results of Entanglement/R\'enyi entropy of wedge holography. 

It should be mentioned that the A-type central charge (\ref{chargegenereld}) has already been presented in equation (2.48) of \cite{Akal:2020wfl} . Since we consider more general solutions here, in addition to the A-type central charge, we can also obtain B-type central charges such as c of (\ref{charge4d}) in this paper.  

Let us end this subsection with a holographic proof of the c-theorem (\ref{atheorem}) for wedge holography.  From the null energy condition on the brane, \cite{Fujita:2011fp} find that $\rho$ is a monotonically decreasing function under the RG flow.  There is a natural geometric interpretation of the monotonicity of $\rho$. See figure 2.   Note that $\rho$ is related to the angle between the brane $Q_1$ and AdS boundary M as  \cite{Takayanagi:2011zk}
\begin{eqnarray}\label{angle}
\phi=\arctan \text{csch}(\rho ).
\end{eqnarray}
As a result, the smaller the tension $\rho$ is, the larger the angle $\phi$ is, and the deeper the brane bends into the bulk. Recall that the AdS boundary M corresponds to UV, while the deep bulk N corresponds to IR. It is clear that  $\rho$ decreases under RG flows. Since $\partial_{\rho} A\ge 0$ from  (\ref{chargegenereld}) , the central charge $A$ also decreases under RG flows. Thus the c-theorem (\ref{atheorem}) is indeed obeyed in wedge holography. Note that, in the viewpoint of BCFT, (\ref{atheorem})  is the g-theorem for boundary central charges. It is interesting that the holographic g-theorem \cite{Fujita:2011fp} for BCFTs becomes holographic c-theorem \cite{Myers:2010tj} for CFTs in the framework of wedge holography. 

\begin{figure}[t]
\centering
\includegraphics[width=10cm]{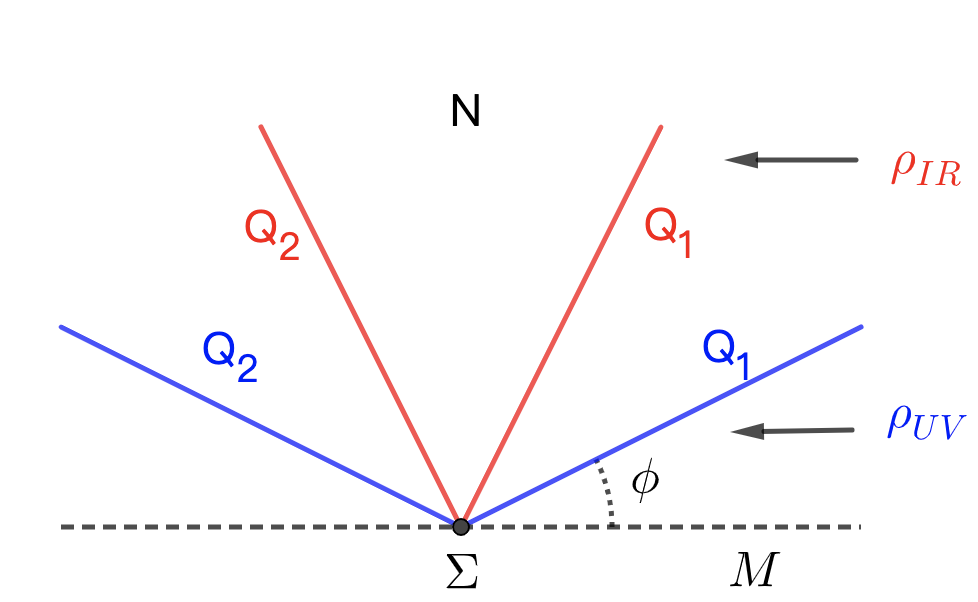}
\caption{UV Planck brane (blue) and  IR Planck brane (red). The smaller the tension $T=(d-1) \tanh\rho$ is, the deeper  the brane bends into the bulk (IR). Thus $\rho$ decreases under RG flows. }
\label{UVIR}
\end{figure}

\subsection{Holographic R\'enyi entropy}

R\'enyi entropy is a complete measure of the quantum entanglement of a subsystem, which is defined by 
\begin{eqnarray}\label{Renyientropy}
S_n=\frac{1}{1-n} \ln \text{tr} \rho_A^n,
\end{eqnarray}
where $n$ is a positive number, $\rho_A=\text{tr}_{\bar{A}}\  \rho$ is the induced density matrix of a subregion $A$. Here $\bar{A}$ denotes the complement of $A$ and $\rho$ is the density matrix of the whole system.  In the limit $n\to 1$, R\'enyi  entropy becomes the von Neumann entropy, which is also called entanglement entropy
\begin{eqnarray}\label{Entanglemententropy}
S_{\text{EE}}=-\text{tr} \rho_A \ln \rho_A.
\end{eqnarray}

\begin{figure}[t]
\centering
\includegraphics[width=10cm]{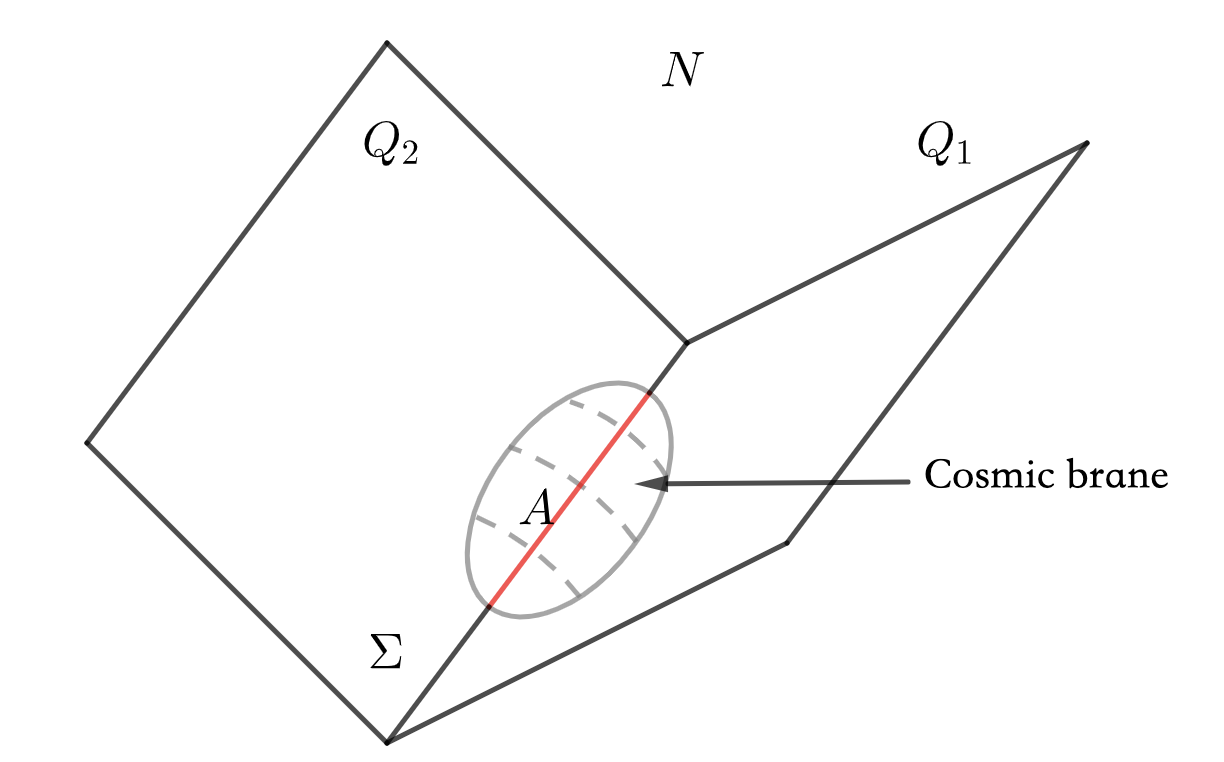}
\caption{Geometry of cosmic brane. The red line denotes a subsystem A on $\Sigma$, and the gray surface denotes the cosmic brane ending on the end-of-world branes $Q_1$ and $Q_2$. Note that the intersection of cosmic brane and $\Sigma$ is $\partial A$. }
\label{cosmicbrane}
\end{figure}

In the gravity dual, R\'enyi entropy can be calculated by the area of cosmic brane \cite{Dong:2016fnf}
\begin{eqnarray}\label{HoloRenyi}
n^2\partial_n\left( \frac{n-1}{n} S_n \right)=\frac{\text{Area(Cosmic Brane}_n)}{4 G_N},
\end{eqnarray}
where the cosmic brane${}_n$ is anchored at the entangling surface $\partial A$.  Notice that since the brane tension \begin{eqnarray}\label{Tn}
T_n=\frac{n-1}{4n G_N}
\end{eqnarray}
is non-zero generally, the cosmic brane backreacts on the bulk geometry.  In the tensionless limit $n\to 1$, the cosmic brane becomes a minimal surface and (\ref{HoloRenyi}) reduces to the Ryu–Takayanagi formula for entanglement entropy \cite{Ryu:2006bv}
\begin{eqnarray}\label{HoloEE}
S_{\text{EE}}=\frac{\text{Area(Minimal Surface)}}{4 G_N}.
\end{eqnarray}

The elegant formula of R\'enyi entropy (\ref{HoloRenyi}) can be straightforwardly generalized to wedge holography. Similar to the case of holographic entanglement entropy \cite{Akal:2020wfl}, the new characteristic in wedge holography is that now the cosmic brane ends on the end-of-world brane $Q_1$ and $Q_2$. 
See figure 3 for example. 
Note that the cosmic brane is codiemsnion two, while the brane $Q_1$ and $Q_2$ are codiemsnion one. Please do not confuse them.  The location of cosmic brane can be fixed by solving  Einstein equations with backreactions  \cite{Dong:2016fnf}. In the case that the solution is not unique, one choose the cosmic brane with the minimal area. 

In principle, by applying (\ref{HoloRenyi}) one can investigate the R\'enyi entropy for general cases. However, the actual challenge is that it is difficult to solve Einstein equations with non-trivial backreactions due to the cosmic brane. So far the known solutions are limited. The hyperbolic black hole is an exact solution, when the entangling surface is a sphere \cite{Hung:2011nu}. There are also perturbation solutions around the hyperbolic black hole \cite{Dong:2016wcf,Bianchi:2016xvf,Chu:2016tps}.  Inspired by \cite{Hung:2011nu}, we choose the following bulk metric for wedge holography
\begin{eqnarray}\label{metricRenyi}
ds^2=dx^2+\cosh^2(x)\left( \frac{dr^2}{f(r)} +f(r) d\tau^2+r^2 d H_{d-2}^2  \right),
\end{eqnarray}
where 
\begin{eqnarray}\label{metricRenyif}
f(r)=r^2-1-\frac{(r_h^2-1)r_h^{d-3}}{r^{d-3}}
\end{eqnarray}
and $d H_{d-2}^2=dr^2+\sinh^2 r d \Omega_{d-3}^2$ is the line element of $(d-2)$-dimensional hyperbolic space with unit curvature.  The R\'enyi index $n$ is related to the temperature of hyperbolic black hole
\begin{eqnarray}\label{metricRenyin}
n=\frac{1}{2\pi T_{tem}}=\frac{2}{f'(r_h)},
\end{eqnarray}
which yields
\begin{eqnarray}\label{metricRenyirh}
r_h=\frac{1+\sqrt{1+\left(d^2-4 d+3\right) n^2}}{(d-1) n}.
\end{eqnarray}
For the solution (\ref{metricRenyi}), the cosmic brane is just the horizon of hyperbolic black hole, whose area is given by
\begin{eqnarray}\label{Renyiarea}
\text{Area(Cosmic Brane}_n)=r_h^{d-2} V_{H_{d-2}}\int_0^{\rho} \cosh^{d-2}(x)dx,
\end{eqnarray}
where $ V_{H_{d-2}}$ is the volume of hyperbolic space.  From (\ref{HoloRenyi},\ref{metricRenyirh},\ref{Renyiarea}), we finally obtain the holographic R\'enyi entropy with sphere entangling surface as
\begin{eqnarray}\label{Renyifinalresult}
S_n=\frac{r_h^{d-2}+r_h^{d}-2 r_h}{ \left(r_h-1\right) \left((d-1) r_h+d-3\right)}\frac{V_{H_{d-2}}}{4 G_N} \int_0^{\rho} \cosh^{d-2}(x)dx.
\end{eqnarray}

Let us discuss briefly the result (\ref{Renyifinalresult}). First, for 2d CFTs, the R\'enyi entropy (\ref{Renyifinalresult}) becomes 
\begin{eqnarray}\label{Renyifinalresult2d}
S_n=\frac{n+1}{ n} \frac{V_{H_{1}}}{8 G_N} \int_0^{\rho} \cosh(x)dx,
\end{eqnarray}
which has the correct n dependence. Second, by taking the limit $n\to 1$, we get the entanglement entropy 
\begin{eqnarray}\label{RenyifinalresultEE}
S_{\text{EE}}=\frac{V_{H_{d-2}}}{4 G_N} \int_0^{\rho} \cosh^{d-2}(x)dx,
\end{eqnarray}
which agrees with the RT formula (\ref{HoloEE}). 
Since $V_{H_{d-2}}$ includes a log term  \cite{Hung:2011nu}
\begin{eqnarray}\label{Vlog}
V_{H_{d-2}} |_{\ln \frac{1}{\epsilon}}=\frac{2\pi^{(d-3)/2}}{\Gamma(\frac{d-1}{2})} (-1)^{(d-3)/2} , 
\end{eqnarray}
(\ref{RenyifinalresultEE}) yields the correct universal term of entanglement entropy (\ref{chargegenereld},\ref{EEfromanomaly}).  This is a double check of our results of holographic Weyl anomaly and holographic R\'enyi entropy,  which is also a support for the wedge holography.

\subsection{Holographic correlation function}

In this subsection, we discuss correlation functions in wedge holography. For simplicity, we focus on the two point functions of stress tensors. We find that the two point functions take the expected form, which can be regarded as a test of wedge holography. 

Following \cite{Liu:1998bu},  we consider the metric fluctuations $H_{ij}$ on the AdS brane $Q_1$
\begin{eqnarray}\label{metricH}
ds^2=dx^2+\cosh^2(x)\frac{dz^2+\delta_{ab}dy^ady^b+H_{ij}dy^{i}dy^{j}}{z^2}
\end{eqnarray}
and choose the gauge
\begin{eqnarray}\label{gaugeH}
H_{zz}(z=0,{\bf{y}})=H_{za}(z=0,{\bf{y}})=0
\end{eqnarray} 
on the corner of the wedge $\Sigma$. Since the contributions from  $Q_1$ and  $Q_2$ are the same, we only need to consider the contributions due to $Q_1$ and double the final results. Here $x^{\mu}=(x,y^i)$ denote coordinates in the bulk N and  $y^i=(z,y^a)$ are coordinates on the brane Q.  
Imposing the Dirichlet boundary condition on $\Sigma$
\begin{eqnarray}\label{BCH}
H_{ab}(z=0,{\bf{y}})=\hat{H}_{ab}({\bf{y}}),
\end{eqnarray} 
we solve the bulk solution
\begin{eqnarray}\label{bulkH}
H_{ij}(z,{\bf{y}})=\frac{\Gamma[d-1](d)}{\pi^{(d-1)/2}\Gamma[(d-1)/2](d-2)}\int d^{d-1}y' \Big{[}  \frac{z^{d-1}}{S^{2(d-1)}}J_{ ia}J_{jb}P_{abcd}\hat{H}_{ad}({\bf{y'}})  \Big]
\end{eqnarray} 
where
\begin{eqnarray}\label{fbf}
&&S^2=z^2+(y_a-y'_a)^2, \nonumber\\
&&P_{abcd}=\frac{1}{2}\left( \delta_{ac} \delta_{bd} +\delta_{ad} \delta_{bc}\right)-\frac{1}{d-1}\delta_{ab}\delta_{ac},\nonumber\\
&&J_{ij}=\delta_{ij}-2\frac{(y_{i}-y'_{i})(y_{j}-y'_{j})}{S^2}.
\end{eqnarray} 

From the renormalized action (\ref{AdSWaction}) plus (\ref{counterterms}), one gets the on-shell quadratic action for $H_{ij}$  \cite{Liu:1998bu},
\begin{eqnarray}\label{I2}
I_2=\frac{1}{16\pi G_N^{(d)}}\int_{\sigma} dy^{d-1} z^{2-d} \left(\frac{1}{4} H_{ab}\partial_z H_{ab} -\frac{1}{2}H_{ab}\partial_b H_{za}\right),
\end{eqnarray} 
where $G_N^{(d)}$ is given by (\ref{Newton's constant}). 
 Substituting (\ref{bulkH}) into (\ref{I2}), we derive
\begin{eqnarray}\label{I2HH}
I_2=\frac{1}{64\pi G_N^{(d)}}\frac{\Gamma[d+1]}{\pi^{(d-1)/2}\Gamma[(d-1)/2](d-2)}\int dy^{d-1} dy'^{d-1} \hat{H}_{ab}({\bf{y}})\Big{[}  \frac{\mathcal{I}_{ab,cd}}{s^{2(d-1)}}  \Big{]}\hat{H}_{cd}({\bf{y'}}),
\end{eqnarray}
where 
\begin{eqnarray}
&&s^2=(y_a-y'_a)^2, \label{s} \\
&&\mathcal{I}_{ab,cd}=\lim_{z\to 0}\frac{1}{2}\left( J_{ac} J_{bd} +J_{ad} J_{ab}\right)-\frac{1}{d-1}\delta_{ab}\delta_{cd},\label{Iijkl}\end{eqnarray} 
From (\ref{I2HH}), we finally obtain the holographic two point function of stress tensor for $\text{CFT}_{d-1}$
\begin{eqnarray}\label{TTfromHH}
<T_{ab}({\bf{y}})T_{cd}({\bf{y'}})>=C_T  \frac{\mathcal{I}_{ab,cd}}{s^{2(d-1)}} ,
\end{eqnarray}
with the central charge
\begin{eqnarray}\label{CTHH}
C_T=\frac{2\Gamma[d+1]}{\pi^{(d-1)/2}\Gamma[(d-1)/2](d-2)} \frac{1}{16\pi G_N}\int_0^{\rho} \cosh^{d-2}(x)dx.
\end{eqnarray}
 Recall that the total central charge is double of the above $C_T$. It is clear that the two point function (\ref{TTfromHH}) takes the correct form. This is a strong support of wedge holography.

\section{More general solutions}

As mentioned in above sections, although it is novel and powerful, the metric (\ref{keysolution}) is not the most general solution to vacuum Einstein equation with NBC (\ref{NBC}) in $d+1$ dimensions. In this section, we discuss the property of general solutions and find a perturbation solution beyond the novel class of solution  (\ref{keysolution}) . We argue that the wedge holography with general solutions is equivalent to AdS/CFT with suitable matter fields. A natural origination of the matter fields are Kaluza-Klein modes from the dimensional reduction.

\subsection{General spacetime on brane}

Let us first discuss the most general spacetime allowed on the brane $Q$. For simplicity, we focus on vacuum Einstein gravity (\ref{action}) without matters in the bulk $N$. The generalization to the theory including matter fields is straightforward.

The induced metric $h_{ij}$ on the brane should obey the Momentum constraint
\begin{eqnarray}\label{MomentumConstraint}
D_i (K^{ij}-K h^{ij})=0,
\end{eqnarray}
and Hamiltonian constraint
\begin{eqnarray}\label{HamiltonianConstraint}
R_h+K^{ij}K_{ij}-K^2+d(d+1)=0,
\end{eqnarray}
where $D_i$ is the covariant derivative on the brane.  Imposing the NBC (\ref{NBC}), i.e., $K_{ij}=\frac{T}{d-1} h_{ij}$, 
we find that the Momentum constraint (\ref{MomentumConstraint}) is satisfied automatically, and the Hamiltonian constraint (\ref{HamiltonianConstraint}) yields
\begin{eqnarray}\label{ConstraintRh}
R_h=\frac{d}{d-1} \left( T^2-(d-1)^2 \right).
\end{eqnarray}
This is the only constraint of the spacetime on the brane. It is remarkable that it is the spacetime with a constant Ricci scalar.  Thus the novel solution (\ref{keysolution}) obeying Einstein equations is a special case of it. In general, there are three kinds of spacetime allowed on the brane 
\begin{eqnarray}\label{threeR}
\begin{cases}
R_h <0 , \   \ \ \text{if}\ \ |T| < (d-1),\\
R_h>0, \  \ \   \text{if}\ \ |T| > (d-1),\\
R_h=0 \  \ \ \ \text{if}\ \ |T| = (d-1).
\end{cases}
\end{eqnarray}
In this section, we focus on the first case, in particular, the asymptotically AdS on the brane. We leave the discussions of the other two kinds of spacetimes to the next section. 

\subsection{Perturbation solution}

It is difficult to solve Einstein equations with the NBC (\ref{NBC}) on both branes $Q_1$ and $Q_2$.  To gain some insight of the solutions, let us first study a simpler case, the perturbation solution near one of the brane.  At the end of this section, we give a perturbation solution which satisfies NBC (\ref{NBC}) on both branes. 

In Gauss normal coordinates, the metric near the brane $Q_1$ takes the following form
\begin{eqnarray}\label{dsnearbrane}
ds^2=d\bar{x}^2+\left( (1+2 \bar{x} \tanh\rho) h_{ij} +\sum_{n=2}^{\infty} \bar{x}^n h^{(n)}_{ij}(y) \right) dy^i dy^j,
\end{eqnarray}
 where $h_{ij}$ are the induced metric on $Q_1$ located at $\bar{x}=0$ and $\bar{x}=x-\rho$ denotes the distance to the brane $Q_1$. Note that $h_{ij}$ of (\ref{dsnearbrane}) is equal to $\cosh^2 (\rho) \bar{h}_{ij}$ of sect. 2.1. Note also that the above $O(x)$ term of the metric is chosen carefully so that it obeys NBC (\ref{NBC}).  Solving Einstein equations (\ref{EOMg}) at the leading order of $x$, we derive 
 \begin{eqnarray}\label{h2ij}
h^{(2)}_{ij}=R_{h\ ij}+\left( d+(2-d) \tanh^2\rho  \right) h_{ij},
\end{eqnarray}
provided the constraint (\ref{ConstraintRh}) is satisfied.  Similarly, we can solve $h^{(n)}_{ij}$ order by order for every given $h_{ij}$ obeying the constraint (\ref{ConstraintRh}). Let us rewrite (\ref{h2ij}) into a more enlightening form
 \begin{eqnarray}\label{Einsteinmatter}
R_{h\ ij}-\frac{R_h}{2}h_{ij}-\frac{(d-1)(d-2)}{2 \cosh^2\rho}h_{ij}=8 \pi G^{(d)}_N T_{ij},
\end{eqnarray}
where $h^{(2)}_{ij}=(1+\tanh ^2(\rho ))h_{ij}+8 \pi G^{(d)}_N T_{ij}$ and $T_{ij}=0$ for the novel solution (\ref{keysolution}). 
Recall that  $h_{ij}=\cosh^2 (\rho) \bar{h}_{ij}$.
We suggest that $T_{ij}$ are the effective matter stress tensors on the brane for the most general solutions. In other words, the special solution (\ref{keysolution}) corresponds to vacuum Einstein gravity on the brane, while the most general solution corresponds to Einstein gravity coupled with matters. It is interesting that the constraint (\ref{ConstraintRh}) yields
 \begin{eqnarray}\label{matterTij}
T^i_{\ i}=0,
\end{eqnarray}
which implies that the effective matter fields are CFTs.  This is consistent with the proposal of doubly holographic models \cite{Bousso:2020kmy,Penington:2019npb,Almheiri:2019psf,Almheiri:2019hni}.  Note that the effective matter fields can only be regarded as CFTs approximately. That is because, in general, the Kaluza-Klein modes from the dimensional reduction are massive. 

The above solutions apply to the region near to one of the brane. Now let us go on to discuss the general solution which works well in the whole region.  For simplicity, let us focus on four dimensions. We apply the method first developed by  \cite{Miao:2017aba} for AdS/BCFT. The main challenge for wedge holography is that the solution should satisfy NBC on two branes instead of only one.  Inspired by \cite{Miao:2017aba}, we choose the following ansatz of metric
\begin{eqnarray}\label{pergeneralsolution}
ds^2=dx^2+\cosh^2(x)\frac{\left(1+ \epsilon\ z^3 f_z(x)\right)dz^2+\left(1+ \epsilon\ z^3 f_y(x)\right)(dy_1^2+dy_2^2)}{z^2} +O(\epsilon)^2,
\end{eqnarray}
and embedding functions of $Q_1$ and $Q_2$
\begin{eqnarray}\label{embedding}
x=\pm (\rho+ \lambda\ \epsilon z^3),
\end{eqnarray}
where $\epsilon$ denote a small perturbation parameter and $\lambda$ is a constant to be determined. We require $f_z(x), f_y(x)$ to be even functions of $x$. As a result, if the brane $Q_1$ obeys NBC (\ref{NBC}), so does the brane $Q_2$.  Substituting (\ref{pergeneralsolution},\ref{embedding}) into Einstein equations (\ref{EOMg}) and imposing the NBC (\ref{NBC}), we solve 
\begin{eqnarray}\label{fzfy}
&&f_z(x)=-2 f_y(x)=\text{sech}(x)+2 \tanh (x) \tan ^{-1}\left(\tanh \left(\frac{x}{2}\right)\right), \\
&&\lambda=\frac{1}{8} \tan ^{-1}\left(\tanh \left(\frac{\rho }{2}\right)\right).
\end{eqnarray}
From the above solution, we obtain the induced metric $h_{ij}$ on both branes
\begin{eqnarray}\label{perinducedhij}
ds_Q^2=\cosh^2(\rho)\frac{\left(1+ \epsilon\ z^3 h_z(\rho)\right)dz^2+\left(1+ \epsilon\ z^3 h_y(\rho)\right)(dy_1^2+dy_2^2)}{z^2} +O(\epsilon)^2,
\end{eqnarray}
where 
\begin{eqnarray}\label{perinducedhijhz}
&&h_z(\rho)=\text{sech}(\rho )+\frac{9}{4} \tanh (\rho ) \tan ^{-1}\left(\tanh \left(\frac{\rho }{2}\right)\right),\\ \label{perinducedhijhy}
&&h_y(\rho)=-\frac{1}{4} \text{sech}(\rho ) \left(2+3 \sinh (\rho ) \tan ^{-1}\left(\tanh \left(\frac{\rho }{2}\right)\right)\right).
\end{eqnarray}
Substituting the induced metric (\ref{perinducedhij}) into (\ref{Einsteinmatter}), we get the effective stress tensors
\begin{eqnarray}\label{perTijmatter}
T^i_{\ j}=\frac{ \epsilon\ z^3 \text{sech}^3(\rho )}{32\pi G^{(d)}_N} \ \text{diag} \left(2,- 1,- 1 \right) +O(\epsilon)^2,
\end{eqnarray}
which are traceless.

As we have shown above and in sect. 4.1, the induced metrics on the brane need not to satisfy the vacuum Einstein equations (\ref{EOMg}). Instead, the only the constraint is that the Ricci scalar is a constant (\ref{ConstraintRh}). This means that the solution space for wedge holography is larger than the one for AdS/CFT with vacuum Einstein equations.  On the other hand, we have showed that the CFTs in wedge holography and AdS/CFT have the same central charges, R\'enyi entropy and correlation functions in sect. 2 and sect. 3. In other words, they must be the same kind of CFTs (up to some background fields) \footnote{Note that the central charges are universal characteristics of CFTs, which are independent of gravitational solutions in the dual theory. Thus the novel solution (\ref{keysolution}) and the more general solution must be dual to the same kind of CFTs with the same central charges. Let us consider a similar case in AdS/CFT. Of course, AdS and AdS black hole are dual to the same kind of CFTs. The only difference is that AdS corresponds to the vacuum state while AdS black hole corresponds to the thermal state of CFTs.}.  To resolve the `contradiction', a natural guess would be that the wedge holography is equivalent to AdS/CFT with suitable ``matter fields "
turned on.  This is reasonable, since after the dimensional reduction along the $x$ direction, there are Kaluza-Klein modes on the brane, which can be regarded as effective matter fields.  For example, the original 5d Kaluza-Klein theory
consists of a massless graviton, a massless vector, a massless scalar, and a tower of charged massive gravitons.  Apart from the massless graviton, we can regard the other compositions as effective ``matter fields''. Another quick guess for the effective theory is the higher derivative gravity. However, this is not a good candidate for reasons below. First, the higher derivative gravity is usually non-unitary, unless the higher derivative terms are chosen carefully.  Second, the effective theory should yield the same Weyl anomaly as wedge holography. However, this is not the case for higher derivative gravity generally. Thus the effective theory is not likely to be a higher derivative gravity. Even if there were higher curvature terms, similar to the case of $f(R)$ gravity they can be rewritten as matter fields formally.
 
For the reasons above, we propose that the wedge holography with vacuum Einstein gravity in $d+1$ dimensions is equivalent to AdS/CFT with suitable ``matter fields'' coupled to gravity in $d$ dimensions generally. It is interesting to work out exactly the effective theory on the branes.  We leave a careful study of this problem to future works.

\subsection{Information from Weyl anomaly}

In this subsection, we discuss the general expression of Weyl anomaly for wedge holography. For simplicity, we take $\text{AdSW}_4/\text{CFT}_2$ as an example below. 

According to \cite{Akal:2020wfl}, the codimension two holography $\text{AdSW}_4/\text{CFT}_2$  can be regarded as a special limit of $\text{AdS}_4/\text{BCFT}_3$. See figure 1 (right). By taking this limit, we can obtain the Weyl anomaly of  $\text{AdSW}_4/\text{CFT}_2$  from that of $\text{AdS}_4/\text{BCFT}_3$ \cite{Miao:2017aba}
\begin{eqnarray}\label{anomaly2dfromBCFT}
&&\mathcal{A}= \frac{1}{16\pi G_N} \int_{\Sigma} dx^{2}\sqrt{\sigma}\Big(  \sinh(\rho) R_{\Sigma}  + \frac{1}{\theta}\  \bar{k}_{ab}\bar{k}^{ab} \Big),
\end{eqnarray}
where $\bar{k}_{ab}$ is the traceless part of extrinsic curvature for BCFTs, and 
$\theta=\frac{\pi}{2}+2\text{tan}{}^{-1}  (\tanh \frac{\rho}{2} )$ is the angle between the brane $Q_2$ and AdS boundary $M$  \cite{Miao:2017aba}.

Some comments are in order. {\bf 1} In the viewpoint of $\text{CFT}_2$, the second term of Weyl anomaly (\ref{anomaly2dfromBCFT}) can be regarded as the contribution from some kinds of background fields. Similar to the gravitational field, the background matter fields such as vector fields and scalar fields can also produce contributions to Weyl anomaly. Take Dirac field with Yukawa couplings as an example
\begin{eqnarray} \label{Fermiaction}
  I=\int_{\Sigma} \sqrt{\sigma} \left( \bar{\psi}i \gamma^{i}\nabla_{i}\psi+
  \phi \bar{\psi} \psi\right).
\end{eqnarray}
By applying heat-kernel method \cite{Vassilevich:2003xt}, we obtain the Weyl anomaly in two dimensions
\begin{eqnarray}\label{anomaly2dDirac}
&&\mathcal{A}= \frac{1}{24\pi} \int_{\Sigma} dx^{2}\sqrt{\sigma}\Big(  R_{\Sigma}  + 12 \phi^2 \Big),
\end{eqnarray}
which takes a similar form as (\ref{anomaly2dfromBCFT}). In particular, the second terms of (\ref{anomaly2dfromBCFT}) and (\ref{anomaly2dDirac}) are both positive. The Weyl anomaly for the theory (\ref{Fermiaction}) in four dimensions can be found in \cite{Chu:2020gwq}.  {\bf 2} In the viewpoint of $\text{AdSW}_4$, the information of $\bar{k}_{ab}$ is encoded in the bulk metric $g_{\mu\nu}$, or equivalently, the brane metric $h_{ij}$. {\bf 3} Comparing (\ref{anomaly2dfromBCFT}) with (\ref{Weylanomaly2d}) of sect.3.1, we notice that only the solutions different from (\ref{keysolution}) could produce the second term of Weyl anomaly (\ref{anomaly2dfromBCFT}) .   {\bf 4} For even-dimensional $\text{CFT}_{2p}$, there are non-trivial contributions to Weyl anomaly such as $ \text{tr} \bar{k}^{2p}$. While for odd-dimensional $\text{CFT}_{2p+1}$, the Weyl anomaly vanish as expected. The potential contribution such as $ \text{tr} \bar{k}^{2p+1}$ cancel, since the extrinsic curvature $k_{ab}$ differs by a minus sign on $Q_1\cap \Sigma$ and $Q_2 \cap \Sigma$ in order to have a well-defined limit from AdS/BCFT.  {\bf 5}  The above comments support the proposal that  $\text{AdSW}_{d+1}/\text{CFT}_{d-1}$ is equivalent to $\text{AdS}_d/\text{CFT}_{d-1}$ with suitable matter fields generally. These matter fields on the brane would yield the non-trivial $ \text{tr} \bar{k}^{2p}$ terms in holographic Weyl anomaly. 

As a summary, the most general solutions of wedge holography should obey the constraints (\ref{ConstraintRh}, \ref{h2ij}) and should produce the general Weyl anomaly (\ref{anomaly2dfromBCFT}).

\section{Generalization to dS/CFT and flat space holography}

 The wedge holography proposed in \cite{Akal:2020wfl} assume that the spacetime on the brane is AdS, which is equivalent to AdS/CFT as we showed above.  In fact, the spacetime on the brane can also be dS and Minkowski spacetime \cite{Karch:2000ct}. We study these cases in this section, and show that they are equivalent to dS/CFT \cite{Strominger:2001pn} and flat space holography \cite{Guica:2008mu,Castro:2010fd}, respectively.

Inspired by (\ref{keysolution}), we choose the following ansatz of the metric
\begin{eqnarray}\label{generalansatz}
ds^2=g_{\mu\nu}dx^{\mu}dx^{\nu}=dx^2+f(x) \bar{h}_{ij}(y) dy^i dy^j.
\end{eqnarray}
Substituting (\ref{generalansatz}) into the normal component of Einstein equations, i.e., $R_{xx}=-d$, we get
\begin{eqnarray}\label{normalEOM}
\frac{f''(x)}{2 f(x)}-\frac{1}{4} \left(\frac{f'(x)}{f(x)}\right)^2=1,
\end{eqnarray}
which can be solved as
\begin{eqnarray}\label{normalsolution}
f(x)=(c_1 e^x+ c_2 e^{-x})^2,
\end{eqnarray}
where $c_1$ and $c_2$ are integral constants. There are three kinds of $f(x)$, which correspond to three types of spacetimes on the brane. Let us summarize them below.
\begin{eqnarray}\label{threef}
f(x)=\begin{cases}
\cosh^2(x), \ \ \ \ \  |T| < (d-1),
\ \ \ \ \ \ \text{asymptotically AdS}\\
\sinh^2(x),  \ \ \ \ \  |T| > (d-1),
\ \ \ \ \ \ \text{asymptotically dS} \\
e^{\pm 2x}, \ \ \ \ \ \ \ \  \ \ |T| = (d-1), \ \ \ \ \ \ \text{asymptotically flat}
\end{cases}
\end{eqnarray}
From (\ref{threeR},\ref{threef}), we learn that different value of tension yields different types of spacetime on the brane. $|T| = (d-1)$ is the critical point, where interesting phase transition could occur. 
It should be mentioned that the metrics with (\ref{threef}) have been discussed in \cite{Karch:2000ct}.   \cite{Karch:2000ct} study only the cases that $h_{ij}$ are the metrics of AdS, dS and Minkowski spacetimes.   Here we find that actually $h_{ij}$ can be relaxed to be any metric obeying Einstein equations in d dimensions.  The case of asymptotically AdS has been investigated in above sections. Let us go on to study the cases for asymptotically dS and asymptotically flat spacetime below.

\subsection{Equivalence to dS/CFT}

Unlike AdS/CFT, dS/CFT is more subtle. For instance, the CFT dual to dS is non-unitary \cite{Strominger:2001pn,Maldacena:2002vr}. In this paper, we do not aim to resolve this non-trivial problem. Instead, we just show that AdS/CFT and dS/CFT can be formally unified in the framework of codimension two holography in asymptotically AdS.  It is widely believed that the holography defined in AdS is well-behaved, thus codimension two holography may shed some light on dS/CFT.  We hope we could address this problem in the future. 

Now let us discuss codimension two holography with the dS brane. Since the discussions are similar to the AdS brane of sect. 2, we do not repeat the calculations below. Instead, we just list the main results. Let us start with some theorems.

{\bf Theorem III} \\
The metric (\ref{generalansatz}) with $f=\sinh^2(x)$ is a solution to Einstein equation (\ref{EOMg}) with a negative cosmological constant in $d+1$ dimensions, 
provided that $\bar{h}_{ij}$ obey Einstein equation with a positive cosmological constant in d dimensions
\begin{eqnarray}\label{EOMhdS}
R_{\bar{h}\ ij}-\frac{R_{\bar{h}}}{2}\bar{h}_{ij}=-\frac{(d-1)(d-2)}{2} \bar{h}_{ij}.
\end{eqnarray}
Here we have set the dS radius $L=1$ for simplicity.

{\bf Theorem IV} \\
The metric (\ref{generalansatz}) with $f=\sinh^2(x)$ satisfies NBC (\ref{NBC}) with $T=(d-1) \coth\rho$ on the branes at $x=\pm \rho$.

Note that the dS brane tension $T=(d-1) \coth\rho$ is larger than the AdS brane tension  $T=(d-1) \tanh\rho$.

The theorem III and theorem IV claim that the metric (\ref{generalansatz}) with $f=\sinh^2(x)$ is a solution to AdS/BCFT with range $-\infty \le x \le \rho$ and a solution to condimension two holography with range $-\rho \le x \le \rho$. 

Similar to the case of AdS brane, we can prove that the codimension two holography $\text{AdSW}_{d+1}/\text{CFT}_{d-1}$ with asymptotically dS space on the brane is equivalent to $\text{dS}_d/\text{CFT}_{d-1}$. Following the approach of sect. 2, we obtain the gravitational action with an asymptotically dS brane as
\begin{eqnarray}\label{dSWaction}
\bar{I}_{\text{AdSW}_{d+1}}
&=&\frac{1}{16\pi G_N}\int_0^{\rho} \sinh^{d}(x)dx \int_{Q_1} \sqrt{\bar{h}} \Big{(} R_{\bar{h}} \text{csch}^2(x)- d\left( 2+(d-1) \coth^2x\right)+d(d-1) \Big{)}\nonumber\\
&+&\frac{1}{8\pi G_N}\int_{Q_1} \sqrt{\bar{h}} \sinh^d\rho \coth\rho \nonumber\\
&=& \frac{1}{16\pi G^{(d)}_N}\int_{Q_1} \sqrt{\bar{h}} \Big{(} R_{\bar{h}} -(d-1)(d-2) \Big{)}=I_{\text{dS}_d}\ ,
\end{eqnarray}
which is equal to the gravitational action $I_{\text{dS}_d}$ for asymptotically dS provided that the Newton's constants are related by
\begin{eqnarray}\label{Newton's constant}
\frac{1}{G^{(d)}_N}=\frac{1}{ G_N}\int_0^{\rho} \sinh^{d-2}(x)dx.
\end{eqnarray}
Similar to the case of AdS brane, one can perform holographic renormalization to make finite the gravitational action. Now we finish the proof of the equivalence.  Following the approaches of sect. 3, we can derive holographic Weyl anomaly, holographic R\'enyi entropy, correlation functions and so on.

\subsection{Equivalence to flat space holography}

Let us go on to generalize our discussions to the flat space holography \cite{Guica:2008mu,Castro:2010fd}. Similarly, we have the following two theorems.

{\bf Theorem V} \\
The metric (\ref{generalansatz}) with $f=\exp(\pm 2x)$ is a solution to Einstein equation (\ref{EOMg}) with a negative cosmological constant in $d+1$ dimensions,
provided that $\bar{h}_{ij}$ obey vacuum Einstein equation in d dimensions
\begin{eqnarray}\label{EOMhflat}
R_{\bar{h}\ ij}-\frac{R_{\bar{h}}}{2}\bar{h}_{ij}=0.
\end{eqnarray}

{\bf Theorem VI} \\
The metric (\ref{generalansatz}) with $f=\exp(\pm 2x)$ satisfies NBC (\ref{NBC}) with $|T|=(d-1)$ on the branes at $x=\text{constant}$.

The above two theorems shows that the metric (\ref{generalansatz}) with $f=\exp(\pm 2x)$ is a solution to AdS/BCFT and codimension two holography. The codimension two holography with flat branes can be obtained as suitable limit $\rho\to\infty$ from those with AdS brane and dS brane. In the large $\rho$ limit, the tension of three kinds of branes coincide
\begin{eqnarray}\label{Tlimit}
\lim_{\rho\to \infty} |T_{\text{AdS}}|=  |T_{\text{AdS}}|= |T_{\text{Min}}|=d-1.
\end{eqnarray}
One can prove the equivalence between codimension two holography with flat branes and flat space holography by taking the limit $\rho\to \infty$ and $L\to \infty$ for AdS brane action (\ref{AdSWaction}) and  dS brane action (\ref{dSWaction}). Recall that we have set $L=1$ in  (\ref{AdSWaction}) and  (\ref{dSWaction}). Recover L, $(d-1)(d-2)$ of  (\ref{AdSWaction}) and  (\ref{dSWaction}) should be replaced by $(d-1)(d-2)/L^2$ which vanish in the large $L$ limit. Thus the AdS brane action (\ref{AdSWaction}) and  dS brane action (\ref{dSWaction}) indeed become the flat brane action in the limit $\rho\to \infty$ and $L\to \infty$.  However, the Newton's constant $G^{(d)}_{N}$ vanish in such limit. 

To get non-zero Newton's constant, let us consider a different case. We choose  $f=\exp(2x)$ with $-\infty\le x\le \rho$, or   $f=\exp(-2x)$ with $ -\rho \le x\le \infty$. It means that we put one of the brane to infinity. In fact, we have a large freedom to choose the location of brane. It is not necessary to set them at $x=\pm \rho$. Without loss of generality, let us take the first choice as an example. We get the gravitational action
\begin{eqnarray}\label{MinWaction}
\tilde{I}_{\text{AdSW}_{d+1}}
&=&\frac{1}{16\pi G_N}\int_{-\infty}^{\rho} \exp(d x)dx \int_{Q_1} \sqrt{\bar{h}} \Big{(}  \exp(-2 x)R_{\bar{h}}-2d \Big{)}\nonumber\\
&+&\frac{1}{8\pi G_N}\int_{Q_1} \sqrt{\bar{h}} \exp(d \rho) \nonumber\\
&=& \frac{1}{16\pi G^{(d)}_N}\int_{Q_1} \sqrt{\bar{h}} \Big{(} R_{\bar{h}} \Big{)}=I_{\text{flat}_d}\ ,
\end{eqnarray}
which is equal to the gravitational action $I_{\text{flat}_d}$ for asymptotically flat space provided that the Newton's constants are given by
\begin{eqnarray}\label{Newton's constant}
\frac{1}{G^{(d)}_N}=\frac{1}{ G_N}\int_{-\infty}^{\rho} \exp\left((d-2)x\right)dx=\frac{1}{ G_N} \frac{e^{(d-2) \rho }}{d-2}.
\end{eqnarray}
Now the Newton's constant $G^{(d)}_{N}$ is finite.  

We leave a careful discussion of flat space holography to future work. It is beyond the main purpose of this paper. Formally, now we have unified AdS/CFT, dS/CFT and flat space holography in the framework of codimension two holography in asymptotically AdS.

\section{Conclusions and Discussions}

In this paper, we construct a class of exact gravitational solutions for wedge holography from the ones in AdS/CFT. We prove that the wedge holography with this novel class of solutions is equivalent to AdS/CFT with vacuum Einstein gravity.  Assuming that AdS/CFT is correct, this equivalence can be regarded as a proof of the recently proposed wedge holography, at least in the classical level for gravity, or equivalently, in large N limit for CFTs.  By applying this powerful equivalence, we derive directly holographic Weyl anomaly, holographic R\'enyi entropy and correlation functions for wedge holography. They all take the expected expressions, which is a support of the codimension two holography for wedges. We also discuss the general solutions for wedge holography. We find that the intrinsic Ricci scalar on the brane is always a constant and there are three types of spacetimes (AdS, dS, flat space) depending on the value of brane tension. We argue that the general solutions with asymptotically AdS branes correspond to AdS/CFT with suitable matter fields (Kaluza-Klein modes) coupled to gravity. In particular, the effective matter fields should produce the $\text{tr} \bar{k}^{2p}$ like contributions to holographic Weyl anomaly.  Finally, we generalize our discussions to dS/CFT and flat space holography. It is remarkable that AdS/CFT, dS/CFT and flat space holography can be unified in the framework of codimension two holography in asymptotically AdS. Since the holography in AdS is well-defined, it may shed some light on dS/CFT and flat space holography. 

There are many interesting open problems worth exploring. We just list some of them. 

{\bf 1}. Find solutions different from (\ref{keysolution}) and study their properties in wedge holography. This is the most interesting part of wedge holography, which ``distinguishes'' it from AdS/CFT.

{\bf 2}. Investigate the wedge holography with matter fields both in the bulk and on the brane. In this paper, we mainly focus on vacuum Einstein gravity on both sides. As a result, we can discuss only the correlation functions for stress tensors instead of scalar operators and currents. The generalization to wedge holography with matter fields is an interesting and important problem. See \cite{HuandMiao} for some progresses. 

{\bf 3}. Study the property of black holes in wedge holography and AdS/BCFT \cite{ChuandMiao}. 

{\bf 4}. Generalize wedge holography to higher derivative gravity. 

{\bf 5}. In this paper, we focus on the classical limit of gravity. It is interesting to study the quantum corrections and see if the equivalence between the wedge holography and AdS/CFT still hold. 

{\bf 6}. Apply wedge holography to discuss the information paradox such as Island and the Page curve of Hawking radiations.  

We hope these interesting problems could be addressed in the future.

\section*{Acknowledgements}

We thank C. S. Chu and J. Ren for valuable discussions on black holes of AdS/BCFT. This work is supported by NSFC grant (No. 11905297) and Guangdong Basic and Applied Basic Research Foundation (No.2020A1515010900).

\appendix

\section{Some formulas}

From the metric (\ref{keysolution}), we derive non-zero affine and curvatures as follows
\begin{eqnarray}\label{affine}
\Gamma^x_{ij}=-\tanh(x) g_{ij}, \ \ \Gamma^i_{xj}=\tanh(x) \delta^i_j,\ \ \Gamma^i_{jk}= \Gamma_{\bar{h}}{}^i_{jk},
\end{eqnarray}
\begin{eqnarray}\label{curvature1}
R^x_{\ ixj}=-g_{ij}, \ \  R^i_{\ jkl}=R_{\bar{h}}{}^i_{\ jkl}-\tanh^2x \ (\delta^i_k g_{jl}-\delta^i_l g_{jk}),
\end{eqnarray}
\begin{eqnarray}\label{curvature2}
R_{xx}=-d,\ \ R_{ij}=R_{\bar{h}\ ij}-\left(1+(d-1)\tanh^2 x\right) g_{ij},
\end{eqnarray}
\begin{eqnarray}\label{curvature3}
R=R_{\bar{h}} \text{sech}^2(x)- d\left( 2+(d-1) \tanh^2x\right),
\end{eqnarray}
where $g_{ij}=\cosh^2(x) \bar{h}_{ij}$ and $(\ )_{\bar{h}}$ denotes the quantity defined by the metric $\bar{h}_{ij}$. 

Let us go on to consider the extrinsic curvatures on brane $Q$. The extrinsic curvature is defined by
\begin{eqnarray}\label{extrinsiccurvature}
K_{\mu\nu}=\hat{h}^{\alpha}_{\mu} \hat{h}^{\beta}_{\nu} \nabla_{\alpha} n_{\beta},
\end{eqnarray}
where $\hat{h}^{\alpha}_{\mu}=\delta^{\alpha}_{\mu}-n^{\alpha} n_{\mu}$ are the projection tensors (induced metric) on $Q$ and $n_{\mu}$ is the outpointing normal vector. Recall that we have $-\infty \le x\le \rho$ for AdS/BCFT and $-\rho \le x\le \rho$ for the codimension two holography. Let us first discuss the brane $Q_1$ located at $x=\rho$.  For the metric (\ref{keysolution}), we have 
\begin{eqnarray}\label{inducedmetric1}
n_{\mu}=(1,0,...,0), \ \ \hat{h}^{\alpha}_{\mu}=\text{dig}(0, 1,...,1).
\end{eqnarray}
Substituting the above equations into (\ref{extrinsiccurvature}), we derive
\begin{eqnarray}\label{extrinsiccurvature1}
K_{xx}=K_{xi}=0, \ K_{ij}=\frac{1}{2} (\partial_x g_{ij})|_{x=\rho}=\tanh\rho \ g_{ij}|_{x=\rho}.
\end{eqnarray}
Now let us go on to discuss brane $Q_2$ located at $x=-\rho$. Note that $n_{\mu}$ is the outpointing normal vector. As a result, $n_{\mu}$ on $Q_2$ differs by a minus sign from the one on $Q_1$. We get
\begin{eqnarray}\label{inducedmetric2}
n_{\mu}=(-1,0,...,0), \ \ \hat{h}^{\alpha}_{\mu}=\text{dig}(0, 1,...,1).
\end{eqnarray}
Substituting (\ref{inducedmetric2}) into (\ref{extrinsiccurvature}), we obtain
\begin{eqnarray}\label{extrinsiccurvature2}
K_{xx}=K_{xi}=0, \ K_{ij}=-\frac{1}{2} (\partial_x g_{ij})|_{x=-\rho}=\tanh\rho \ g_{ij}|_{x=-\rho}.
\end{eqnarray}
From (\ref{extrinsiccurvature1}, \ref{extrinsiccurvature2}), we find that the solution (\ref{keysolution}) obeys NBC (\ref{NBC}) on both branes $Q_1$ and $Q_2$. In other words,  (\ref{keysolution}) is a solution to both AdS/BCFT and AdSW/CFT.


\begin{thebibliography}{00}

\bibitem{tHooft:1993dmi}
G.~'t Hooft,
Conf. Proc. C \textbf{930308}, 284-296 (1993)
[arXiv:gr-qc/9310026 [gr-qc]].

\bibitem{Susskind:1994vu}
L.~Susskind,
J. Math. Phys. \textbf{36}, 6377-6396 (1995)
[arXiv:hep-th/9409089 [hep-th]].


\bibitem{Maldacena:1997re}
  J.~M.~Maldacena,
  Int.\ J.\ Theor.\ Phys.\  {\bf 38}, 1113 (1999)
  [Adv.\ Theor.\ Math.\ Phys.\  {\bf 2}, 231 (1998)]
  [hep-th/9711200].
  
\bibitem{Gubser:1998bc}
S.~S.~Gubser, I.~R.~Klebanov and A.~M.~Polyakov,
Phys. Lett. B \textbf{428}, 105-114 (1998)
[arXiv:hep-th/9802109 [hep-th]].

\bibitem{Witten:1998qj}
E.~Witten,
Adv. Theor. Math. Phys. \textbf{2}, 253-291 (1998)
[arXiv:hep-th/9802150 [hep-th]].

\bibitem{Sakai:2004cn}
T.~Sakai and S.~Sugimoto,
Prog. Theor. Phys. \textbf{113} (2005), 843-882
doi:10.1143/PTP.113.843
[arXiv:hep-th/0412141 [hep-th]].


\bibitem{Erlich:2005qh}
J.~Erlich, E.~Katz, D.~T.~Son and M.~A.~Stephanov,
Phys. Rev. Lett. \textbf{95}, 261602 (2005)
doi:10.1103/PhysRevLett.95.261602
[arXiv:hep-ph/0501128 [hep-ph]].

\bibitem{Sakai:2005yt}
T.~Sakai and S.~Sugimoto,
Prog. Theor. Phys. \textbf{114}, 1083-1118 (2005)
doi:10.1143/PTP.114.1083
[arXiv:hep-th/0507073 [hep-th]].


\bibitem{Rangamani:2016dms}
M.~Rangamani and T.~Takayanagi,
Lect. Notes Phys. \textbf{931}, pp.1-246 (2017)
[arXiv:1609.01287 [hep-th]].

\bibitem{Hartnoll:2009sz}
S.~A.~Hartnoll,
Class. Quant. Grav. \textbf{26}, 224002 (2009)
[arXiv:0903.3246 [hep-th]].

\bibitem{Strominger:2001pn}
A.~Strominger,
JHEP \textbf{10}, 034 (2001)
[arXiv:hep-th/0106113 [hep-th]].

\bibitem{Maldacena:2002vr}
J.~M.~Maldacena,
JHEP \textbf{05}, 013 (2003)
[arXiv:astro-ph/0210603 [astro-ph]].

\bibitem{Alishahiha:2004md}
M.~Alishahiha, A.~Karch, E.~Silverstein and D.~Tong,
AIP Conf. Proc. \textbf{743}, no.1, 393-409 (2004)
[arXiv:hep-th/0407125 [hep-th]].

\bibitem{Alishahiha:2005dj}
M.~Alishahiha, A.~Karch and E.~Silverstein,
JHEP \textbf{06}, 028 (2005)
[arXiv:hep-th/0504056 [hep-th]].

\bibitem{Dong:2018cuv}
X.~Dong, E.~Silverstein and G.~Torroba,
JHEP \textbf{07}, 050 (2018)
[arXiv:1804.08623 [hep-th]].

\bibitem{Guica:2008mu}
M.~Guica, T.~Hartman, W.~Song and A.~Strominger,
Phys. Rev. D \textbf{80}, 124008 (2009)
[arXiv:0809.4266 [hep-th]].

\bibitem{Castro:2010fd}
A.~Castro, A.~Maloney and A.~Strominger,
Phys. Rev. D \textbf{82}, 024008 (2010)
[arXiv:1004.0996 [hep-th]].

\bibitem{Bagchi:2010zz}
A.~Bagchi,
Phys. Rev. Lett. \textbf{105}, 171601 (2010)
[arXiv:1006.3354 [hep-th]].

\bibitem{Bagchi:2016bcd}
A.~Bagchi, R.~Basu, A.~Kakkar and A.~Mehra,
JHEP \textbf{12}, 147 (2016)
[arXiv:1609.06203 [hep-th]].


\bibitem{Randall:1999ee}
L.~Randall and R.~Sundrum,
Phys. Rev. Lett. \textbf{83}, 3370-3373 (1999)
[arXiv:hep-ph/9905221 [hep-ph]].

\bibitem{Randall:1999vf}
L.~Randall and R.~Sundrum,
Phys. Rev. Lett. \textbf{83}, 4690-4693 (1999)
[arXiv:hep-th/9906064 [hep-th]].

\bibitem{Karch:2000ct}
A.~Karch and L.~Randall,
JHEP \textbf{05}, 008 (2001)
[arXiv:hep-th/0011156 [hep-th]].

\bibitem{Miyaji:2015yva}
M.~Miyaji and T.~Takayanagi,
PTEP \textbf{2015}, no.7, 073B03 (2015)
[arXiv:1503.03542 [hep-th]].

\bibitem{Takayanagi:2018pml}
T.~Takayanagi,
JHEP \textbf{12}, 048 (2018)
[arXiv:1808.09072 [hep-th]].

\bibitem{Takayanagi:2011zk}
  T.~Takayanagi,
  Phys.\ Rev.\ Lett.\  {\bf 107} (2011) 101602
  [arXiv:1105.5165 [hep-th]].
  
\bibitem{Fujita:2011fp}
M.~Fujita, T.~Takayanagi and E.~Tonni,
JHEP \textbf{11}, 043 (2011)
[arXiv:1108.5152 [hep-th]].

\bibitem{Nozaki:2012qd}
M.~Nozaki, T.~Takayanagi and T.~Ugajin,
JHEP \textbf{06}, 066 (2012)
[arXiv:1205.1573 [hep-th]].

\bibitem{Miao:2018qkc}
R.~X.~Miao,
JHEP \textbf{02}, 025 (2019)
[arXiv:1806.10777 [hep-th]].

\bibitem{Miao:2017gyt}
R.~X.~Miao, C.~S.~Chu and W.~Z.~Guo,
Phys. Rev. D \textbf{96}, no.4, 046005 (2017)
[arXiv:1701.04275 [hep-th]].

\bibitem{Chu:2017aab}
C.~S.~Chu, R.~X.~Miao and W.~Z.~Guo,
JHEP \textbf{04}, 089 (2017)
[arXiv:1701.07202 [hep-th]].


\bibitem{Akal:2020wfl}
I.~Akal, Y.~Kusuki, T.~Takayanagi and Z.~Wei,
[arXiv:2007.06800 [hep-th]].

\bibitem{Bousso:2020kmy}
R.~Bousso and E.~Wildenhain,
[arXiv:2006.16289 [hep-th]].

\bibitem{Penington:2019npb}
G.~Penington,
JHEP \textbf{09}, 002 (2020)
[arXiv:1905.08255 [hep-th]].

\bibitem{Almheiri:2019psf}
A.~Almheiri, N.~Engelhardt, D.~Marolf and H.~Maxfield,
JHEP \textbf{12}, 063 (2019)
[arXiv:1905.08762 [hep-th]].

\bibitem{Almheiri:2019hni}
A.~Almheiri, R.~Mahajan, J.~Maldacena and Y.~Zhao,
JHEP \textbf{03}, 149 (2020)
[arXiv:1908.10996 [hep-th]].


\bibitem{Rozali:2019day}
M.~Rozali, J.~Sully, M.~Van Raamsdonk, C.~Waddell and D.~Wakeham,
JHEP \textbf{05}, 004 (2020)
[arXiv:1910.12836 [hep-th]].

\bibitem{Chen:2019uhq}
H.~Z.~Chen, Z.~Fisher, J.~Hernandez, R.~C.~Myers and S.~M.~Ruan,
JHEP \textbf{03}, 152 (2020)
[arXiv:1911.03402 [hep-th]].

\bibitem{Almheiri:2019psy}
A.~Almheiri, R.~Mahajan and J.~E.~Santos,
SciPost Phys. \textbf{9}, no.1, 001 (2020)
[arXiv:1911.09666 [hep-th]].

\bibitem{Kusuki:2019hcg}
Y.~Kusuki, Y.~Suzuki, T.~Takayanagi and K.~Umemoto,
[arXiv:1912.08423 [hep-th]].

\bibitem{Balasubramanian:2020hfs}
V.~Balasubramanian, A.~Kar, O.~Parrikar, G.~Sárosi and T.~Ugajin,
[arXiv:2003.05448 [hep-th]].

\bibitem{Sully:2020pza}
J.~Sully, M.~Van Raamsdonk and D.~Wakeham,
[arXiv:2004.13088 [hep-th]].

\bibitem{Geng:2020qvw}
H.~Geng and A.~Karch,
[arXiv:2006.02438 [hep-th]].

\bibitem{Chen:2020uac}
H.~Z.~Chen, R.~C.~Myers, D.~Neuenfeld, I.~A.~Reyes and J.~Sandor,
[arXiv:2006.04851 [hep-th]].

\bibitem{Dong:2020uxp}
X.~Dong, X.~L.~Qi, Z.~Shangnan and Z.~Yang,
[arXiv:2007.02987 [hep-th]].

\bibitem{Arias:2019zug}
C.~Arias, F.~Diaz, R.~Olea and P.~Sundell,
JHEP \textbf{04}, 124 (2020)
[arXiv:1906.05310 [hep-th]].

\bibitem{Arias:2019pzy}
C.~Arias, F.~Diaz and P.~Sundell,
Class. Quant. Grav. \textbf{37}, no.1, 015009 (2020)
[arXiv:1901.04554 [hep-th]].

\bibitem{Geng:2020kxh}
H.~Geng,
[arXiv:2005.00021 [hep-th]].


\bibitem{Emparan:1999pm}
R.~Emparan, C.~V.~Johnson and R.~C.~Myers,
Phys. Rev. D \textbf{60}, 104001 (1999)
[arXiv:hep-th/9903238 [hep-th]].

\bibitem{Park:2001jh}
I.~Y.~Park, C.~N.~Pope and A.~Sadrzadeh,
Class. Quant. Grav. \textbf{19}, 6237-6258 (2002)
doi:10.1088/0264-9381/19/23/319
[arXiv:hep-th/0110238 [hep-th]].

  

   \bibitem{ChuandMiao}
   Chong-Sun Chu, Rong-Xin Miao, Black hole and Island in AdS/BCFT, in preparing. 
   
    \bibitem{HuandMiao}
 Peng-Ju Hu, Rong-Xin Miao, Wedge Holography with Matter Fields, in preparing. 
   
   
   
\bibitem{Balasubramanian:1999re}
  V.~Balasubramanian and P.~Kraus,
  Commun.\ Math.\ Phys.\  {\bf 208} (1999) 413
  [hep-th/9902121].


\bibitem{deHaro:2000vlm}
  S.~de Haro, S.~N.~Solodukhin and K.~Skenderis,
  Commun.\ Math.\ Phys.\  {\bf 217} (2001) 595
  [hep-th/0002230].
  
\bibitem{Hayward:1993my}
G.~Hayward,
Phys. Rev. D \textbf{47}, 3275-3280 (1993)
  
\bibitem{Brill:1994mb}
D.~Brill and G.~Hayward,
Phys. Rev. D \textbf{50}, 4914-4919 (1994)
[arXiv:gr-qc/9403018 [gr-qc]].
  
 \bibitem{Weylanomaly} See, for example:
M. J. Duff, “Observations On Conformal Anomalies,” Nucl. Phys. B 125, 334 (1977); M. J. Duff, “Twenty years of the Weyl anomaly,” Class. Quant. Grav. 11, 1387 (1994) [arXiv:hep-th/9308075];
S. Deser, A. Schwimmer, “Geometric classification of conformal anomalies in arbitrary dimensions,” Phys. Lett. B309, 279-284 (1993) [hep-th/9302047].


\bibitem{Hung:2011xb}
L.~Y.~Hung, R.~C.~Myers and M.~Smolkin,
JHEP \textbf{04}, 025 (2011)
[arXiv:1101.5813 [hep-th]].

\bibitem{Cardy:1988cwa}
J.~L.~Cardy,
Phys. Lett. B \textbf{215}, 749-752 (1988)

\bibitem{Zamolodchikov:1986gt}
A.~B.~Zamolodchikov,
JETP Lett. \textbf{43}, 730-732 (1986)

\bibitem{Komargodski:2011vj}
Z.~Komargodski and A.~Schwimmer,
JHEP \textbf{12}, 099 (2011)
[arXiv:1107.3987 [hep-th]].


\bibitem{Myers:2010tj}
R.~C.~Myers and A.~Sinha,
JHEP \textbf{01}, 125 (2011)
[arXiv:1011.5819 [hep-th]].

\bibitem{Osborn:1989td}
H.~Osborn,
Phys. Lett. B \textbf{222}, 97-102 (1989)

\bibitem{Freedman:1999gp}
D.~Z.~Freedman, S.~S.~Gubser, K.~Pilch and N.~P.~Warner,
Adv. Theor. Math. Phys. \textbf{3}, 363-417 (1999)
[arXiv:hep-th/9904017 [hep-th]].

\bibitem{Girardello:1998pd}
L.~Girardello, M.~Petrini, M.~Porrati and A.~Zaffaroni,
JHEP \textbf{12}, 022 (1998)
[arXiv:hep-th/9810126 [hep-th]].

\bibitem{Girardello:1999bd}
L.~Girardello, M.~Petrini, M.~Porrati and A.~Zaffaroni,
Nucl. Phys. B \textbf{569}, 451-469 (2000)
[arXiv:hep-th/9909047 [hep-th]].

\bibitem{Myers:2010xs}
R.~C.~Myers and A.~Sinha,
Phys. Rev. D \textbf{82}, 046006 (2010)
[arXiv:1006.1263 [hep-th]].


\bibitem{Henningson:1998gx}
M.~Henningson and K.~Skenderis,
JHEP \textbf{07}, 023 (1998)
[arXiv:hep-th/9806087 [hep-th]].

\bibitem{Imbimbo:1999bj}
C.~Imbimbo, A.~Schwimmer, S.~Theisen and S.~Yankielowicz,
Class. Quant. Grav. \textbf{17}, 1129-1138 (2000)
[arXiv:hep-th/9910267 [hep-th]].

\bibitem{Miao:2013nfa}
R.~X.~Miao,
Class. Quant. Grav. \textbf{31}, 065009 (2014)
[arXiv:1309.0211 [hep-th]].



\bibitem{Dong:2016fnf}
X.~Dong,
Nature Commun. \textbf{7}, 12472 (2016)
[arXiv:1601.06788 [hep-th]].

\bibitem{Ryu:2006bv}
S.~Ryu and T.~Takayanagi,
Phys. Rev. Lett. \textbf{96}, 181602 (2006)
[arXiv:hep-th/0603001 [hep-th]].

\bibitem{Hung:2011nu}
L.~Y.~Hung, R.~C.~Myers, M.~Smolkin and A.~Yale,
JHEP \textbf{12}, 047 (2011)
[arXiv:1110.1084 [hep-th]].

\bibitem{Dong:2016wcf}
X.~Dong,
Phys. Rev. Lett. \textbf{116}, no.25, 251602 (2016)
[arXiv:1602.08493 [hep-th]].

\bibitem{Bianchi:2016xvf}
L.~Bianchi, S.~Chapman, X.~Dong, D.~A.~Galante, M.~Meineri and R.~C.~Myers,
JHEP \textbf{11}, 180 (2016)
[arXiv:1607.07418 [hep-th]].

\bibitem{Chu:2016tps}
C.~S.~Chu and R.~X.~Miao,
JHEP \textbf{12}, 036 (2016)
[arXiv:1608.00328 [hep-th]].


\bibitem{Liu:1998bu}
H.~Liu and A.~A.~Tseytlin,
Nucl. Phys. B \textbf{533}, 88-108 (1998)
[arXiv:hep-th/9804083 [hep-th]].


\bibitem{Miao:2017aba}
R.~X.~Miao and C.~S.~Chu,
JHEP \textbf{03} (2018), 046
[arXiv:1706.09652 [hep-th]].

\bibitem{Vassilevich:2003xt} 
  D.~V.~Vassilevich,
  Phys.\ Rept.\  {\bf 388}, 279 (2003)

\bibitem{Chu:2020gwq}
C.~S.~Chu and R.~X.~Miao,
JHEP \textbf{08}, 134 (2020)
[arXiv:2005.12975 [hep-th]].





\end{thebibliography}
\end{document}